\documentclass[preprint2]{proto}
\usepackage{times}
\usepackage{xcolor}
\usepackage{graphicx}
\usepackage[nointegrals]{wasysym}
\usepackage{float}
\usepackage{dblfloatfix}
\usepackage{color}
\usepackage{amsmath}
\usepackage{tabularx}
\usepackage[normalem]{ulem}
\graphicspath{{./}{Figures/}}

\newcommand\rev[1]{\textcolor{black}{{\scriptsize} #1}}
\newcommand\fix[1]{\textcolor{black}{{\scriptsize} #1}}
\newcommand\final[1]{\textcolor{black}{{\scriptsize} #1}}

\begin{document}

\title{\textbf{\LARGE RADIATIVE PROCESSES AS DIAGNOSTICS OF COMETARY ATMOSPHERES }}

\author {\textbf{\large D. Bodewits}}
\affil{\small\em Physics Department, Leach Science Center, Auburn University, AL 36832, USA}

\author {\textbf{\large B. P. Bonev}}
\affil{\small\em Department of Physics, American University, Washington, DC 20016, USA}

\author {\textbf{\large M. A. Cordiner}}
\affil{\small\em Astrochemistry Laboratory, NASA Goddard Space Flight Center, 8800 Greenbelt Road, Greenbelt, MD 20771, USA\\ Department of Physics, Catholic University of America, Washington, DC 20064, USA}

\author {\textbf{\large G. L. Villanueva}}
\affil{\small\em Solar System Exploration Division, NASA Goddard Space Flight Center, Greenbelt MD 20771, USA}
\begin{abstract}

\begin{list}{ } {\rightmargin 1in}
\baselineskip = 11pt
\parindent=1pc
{\small 
\noindent \rev{In this chapter}, we provide a review of radiative processes in cometary atmospheres \rev{spanning a broad range of wavelengths, from radio to X-rays. We} focus on spectral modeling, observational opportunities, and anticipated challenges in the interpretation of new observations, based on our current understanding of the atomic and molecular processes occurring in the atmospheres of small, icy bodies. Close to the surface, \rev{comets possess} a thermalized atmosphere that traces the irregular shape of the nucleus. Gravity is too low to retain the gas, which flows out to form a large, collisionless exosphere (coma) \rev{that interacts with the heliospheric radiation environment.} As such, cometary comae represent conditions that are familiar in the context of planetary atmosphere studies. However, the outer comae are tenuous, with densities lower than those found in vacuum chambers on Earth. Comets therefore provide us with unique natural laboratories that can be understood using state-of-the-art theoretical treatments of the relevant microphysical processes. Radiative processes offer direct diagnostics of the local physical conditions, as well as the macroscopic coma properties. \rev{These can be used to improve our understanding of comets and other astrophysical environments such as icy moons and the interstellar medium}.\\~\\~\\~}
\end{list}
\end{abstract} 

\section{\textbf{Introduction}}
\label{sec:intro}
The spectroscopic analysis of cometary comae dates back to \citet{Huggins1868}, who matched the optical spectrum of comet 2P/Encke to the carbon emission signatures observed in flames from an organic gas (see \citealt{Swings1965, Feldman2004} for historical reviews). As shown in Figure 1, the optical spectrum of comets is typically dominated by the emission of several radicals (such as C$_2$, CN, and OH), superimposed on the emission of a blackbody spectrum (sun light reflected by dust in the coma). Since these first observations, the spectroscopic investigation of comets has greatly expanded.

\begin{figure*}[t]
\centering
\includegraphics[width=6in]{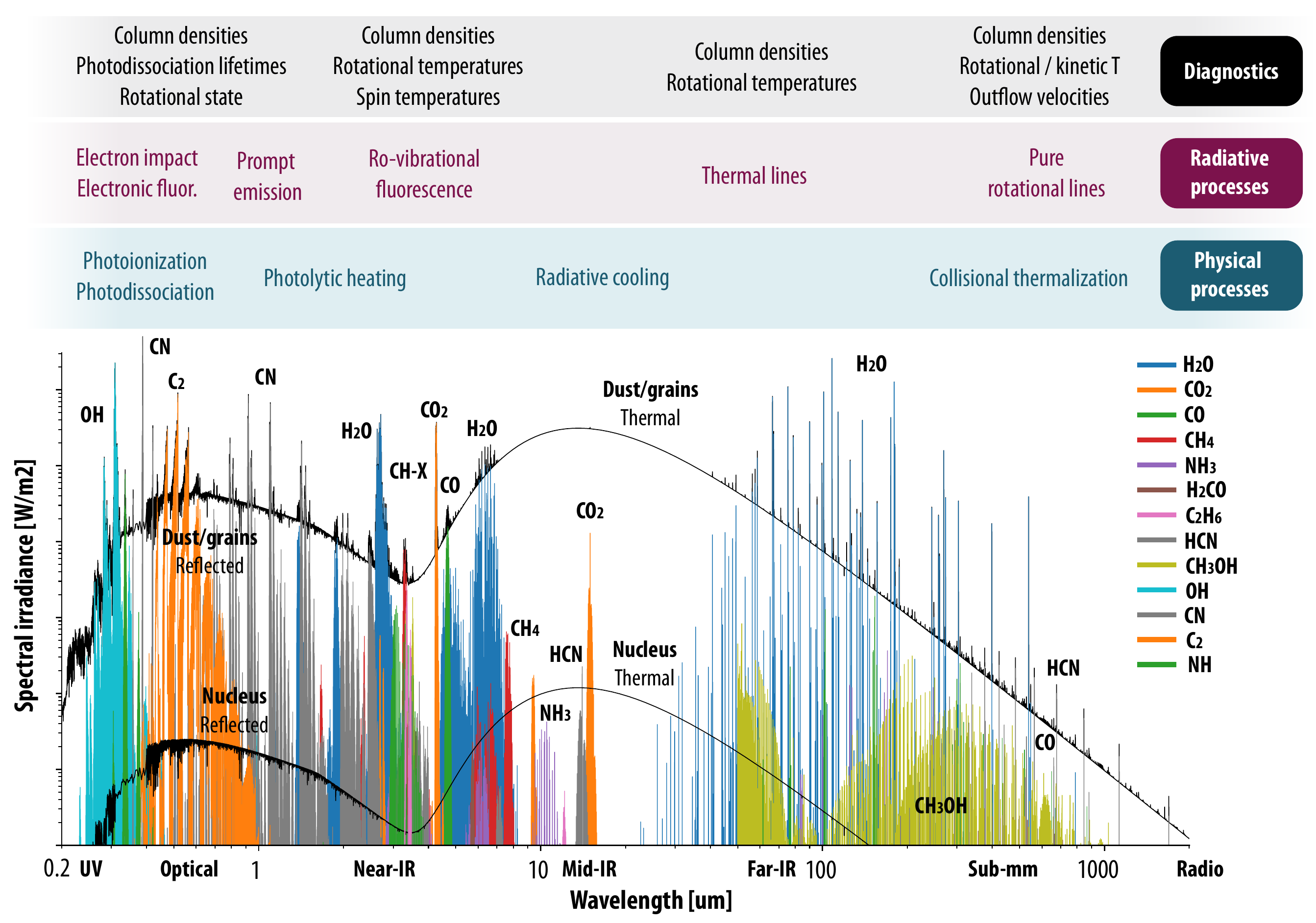}
\caption{Example (synthetic) spectrum of cometary emissions and their relationship to retrievable diagnostics, radiative processes and the associated physical processes.  This synthetic spectrum includes only a subset of cometary species, and was synthesized using the Planetary Spectrum Generator (PSG; \citealt{psg2018}; https://psg.gsfc.nasa.gov), capturing the dominant radiative processes at a resolving power of 5000. The broad cometary continuum at the assumed conditions ($Q=10^{29}$ s$^{-1}$, diameter = 5 km, heliocentric and geocentric distance = 1 au, aperture = $100''$) is usually dominated by dust grains, yet for distant and low activity comets, the nucleus can become the main thermal continuum source. \label{fig:spectral_overview}}
\end{figure*}

\rev {An important driver for these studies is that studying cometary compositions allows us to probe the physics and chemistry of our early solar system \citep{Biver2023}. Observations of coma emission provide measurements that are diagnostic for various properties of both the nucleus and the coma. These measurements include column densities, production rates, relative abundances (also referred to as `mixing ratios'), rotational and kinetic temperatures, photodissociation scale lengths and lifetimes, gas outflow velocities, plasma structures, and the rotational state of the nucleus, among others. This large suite of `metrics' is then interpreted in terms of chemical composition and isotopic abundances, as well as the long-term storage, physico-chemical evolution, and release of volatiles in small proto-bodies, both directly from the nucleus and from secondary sources.} 

\rev{Comets are natural laboratories for studying a wide variety of chemical and physical phenomena owing to the range of processes at work (such as ice sublimation, volatile release and acceleration, thermal processing of grains, photochemistry, ionospheric processes) and their variability over the comet’s orbit about the Sun. Understanding these processes provides insights into the complex interactions between matter and radiation, knowledge of which informs other areas of astrophysics, astrochemistry, and planetary science.} For instance, studies of cometary ice abundances help to understand the impact of radiation on the evolution of primordial ices \citep{Garrod2019}, and help constrain theories for the gas and ice-phase chemistry occurring in interstellar clouds and protoplanetary disks \citep{Eistrup2019, Price2021, Bergin2023}. \rev{The low densities and sublimative outflows of cometary comae offer an accessible analogue to planetary exospheres. With volatile mass loss rates of tens to thousands kg/s, comets can also provide a baseline for the study of outgassing from (and the chemistry of) icy moons, asteroids, and trans-Neptunian objects.} X-ray observations of comets paved the way to remotely study plasma interactions in a multitude of astrophysical environments \citep{Dennerl2010}. Last but not least, comets provide a road map for the detection of volatiles (including molecules that are possibly relevant to the origin of life) in low-density environments within the solar system and beyond \citep{strom2020,cordiner2020,bodewits2020,villanueva2021}.

Cometary comae can span several hundreds of thousands kilometers in size, and their ions tails can stretch to over an astronomical unit in some cases \citep{Neugebauer2007}. Remote sensing is a prime method for studying such large structures, yet interpreting the light emitted by atoms and molecules requires a detailed understanding of the physical processes and how they operate under the specific conditions of cometary atmospheres. \fix{These processes operate in the coma across all energies and wavelengths, but certain wavelengths provide more sensitivity to specific phenomena. For instance, fluorescence typically dominates the infrared/optical emission, but such processes also impact the ground rotational levels sampled at sub-mm/radio wavelengths.Specific wavelength ranges are better suited to probe certain processes in the coma, and Observations at specific wavelengths can thus be tailored to obtain key diagnostics that distinguish and connect physical and chemical processing and primordial properties} (see Figure~\ref{fig:spectral_overview}). This direct link between \rev{the diagnostic radiative emissions and the associated physical processes} also simplifies the interpretation of spectroscopic data, in that only a specific set of processes needs to be included in the modeling and interpretation of the data. 

\begin{figure*}
\centering
\includegraphics[width=6in]{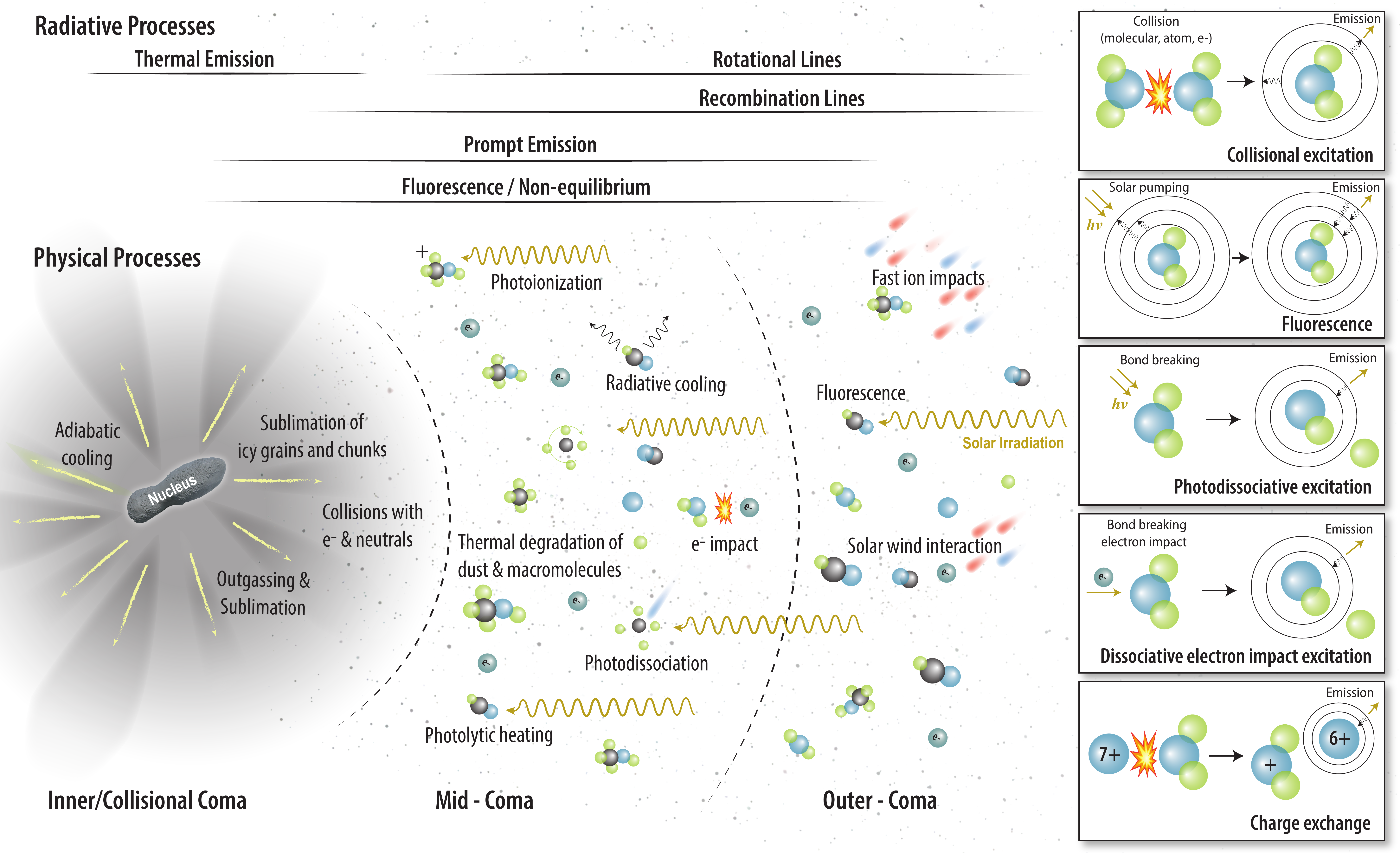}
\caption{Overview of physical regimes and structure of the coma (not to scale), and the dominant physical/chemical processes that factor into the molecular excitation and emission processes considered in this chapter. Key emission processes and their associated energetic driving mechanisms are shown on the right. Horizontal bars at the top indicate the approximate spatial ranges over which these emission processes occur. \label{fig:processes}}
\end{figure*}

A key challenge in interpreting cometary emission is the disequilibrium between the excitation and ambient temperature. This is due to the intrinsically tenuous nature of the atmosphere (because of a lack of equilibrating and thermalizing collisions), combined with a deeply penetrating solar radiation field that sublimates icy grains, breaks molecular bonds (photodissocation), ionizes atoms/molecules (photoionization), and also leads to non-equilibrated emissions via solar pumping (fluorescence).

Low-density regimes (such as the atmospheres of main-belt-comets, Centaurs, distant comets) are challenging to observe, but with the launch of the \emph{James Webb Space Telescope} and many other large telescopes and space observatories planned (or in construction), the observational exploration of comets will soon enter a new era (see Section \ref{sec:nextsection}). These new observational facilities will further demonstrate the limits of our current understanding of the processes that occur in these tenuous regimes. Their interpretation will require more detailed theoretical treatments and advanced models that incorporate accurate laboratory data, molecular constants, and parameterizations obtained under conditions that represent those found in the coma.

\fix{A review of excitation of cometary molecules and the resulting spectra was provided by \citet{boc_comets2} for parent volatiles and by \citet{Feldman2004} for fragment species}. Building on this, in this chapter we discuss the processes that are responsible for the most notable emissions in comets, and which yield diagnostics that allow the remote characterization of physical conditions and chemical parameters of the nucleus and the coma. We organize the discussion by process rather than wavelength, since cometary processes span a broad range of energies and are therefore generally not restricted to a particular wavelength regime. We also explore the limitations of how we currently model and interpret these diagnostics and discuss the challenges and possible steps to be taken to advance the field of coma remote sensing. 

\section{\textbf{Radiative Transfer in the Coma}}
\label{sec:radiative_trans}

\subsection{Physical Structure of the Coma}\label{sec:neutrals}

As ices in the nucleus sublimate, the resulting gases drag icy particles and refractory materials along and form an expanding atmosphere of gas and dust, referred to as the coma (Figure~\ref{fig:processes}). The thermal velocity of the gas molecules exceeds the low escape velocity of comet nuclei (typically $\lesssim0.1$ m\,s$^{-1}$), so instead of forming a hydrostatic atmosphere as in planets, an expanding atmosphere/exosphere is established, the dynamics of which are dominated by outgassing and acceleration processes. The gas and dust decouple within roughly 10~km of the surface \citep{Gerig2018}. As the gas continues to expand, the density drops \rev{rapidly (approximately as ${\rho^{-2}}$, where $\rho$ is the distance from the nucleus)} and the neutral coma begins to depart from thermal and chemical equilibrium. This ultimately results in the formation of a well-defined ionosphere that interacts with the solar wind \rev{(see the chapter by \citealt{Beth2023})}.

The outflow velocity of the expanding coma ($v_{out}$) is governed by the balance of heating and cooling processes \citep{rodgers04}, which vary with coma position, potentially leading to measurable variations in $v_{out}$ with radius. However, for most observational applications (integrating over large spatial apertures), a constant value of $v_{out}$ is assumed. Spatially-averaged gas outflow velocities are commonly derived at radio wavelengths using heterodyne spectroscopy to measure spectrally resolved (Doppler) line profiles \citep{biv99,rot21}. In the optical and infrared, information on the outflow velocity can also be obtained from temporally-resolved imaging of outward-moving coma density structures \citep{schleicher2004,knight21}.

There are multiple different approaches to modeling the density, velocity, and structure of cometary atmospheres \citep{Combi2004,rodgers04}. \citet{haser57} provided an analytical basis for understanding the large-scale chemical structure of cometary comae, under the assumption of an isotropic outflow of gases at a constant radial velocity away from the nucleus. Photodissociation by the solar radiation field leads to exponential decay of the gas radial density on top of the $\rho^{-2}$ spherical expansion (see \citet{has20} for English translation). \fix{For parent species, volume densities following Haser are described as N($\rho$) = $\frac{Q}{4\pi\rho^2v_{out}} e^\frac{-\rho\beta}{v_{out}}$, where $Q$ is the production rate, $\rho$ is the distance to the nucleus, $v_{{out}}$ is the expansion velocity and $\beta$ is the photodissociation rate.}

Due to its simplicity and broad applicability, the Haser model continues to be the primary method used for \rev{linking column densities of atoms and molecules to the production rates}. However, the Haser model has several limitations. It is time-independent, with the assumption of uniform, isotropic outgassing, and \rev{it cannot reproduce observed column density profiles with experimentally determined lifetimes of fragment species in the solar radiation field} \citep{Combi1980}. To work around this, studies of daughter fragments adopt empirical scale lengths \citep{Cochran1993,AHearn1995} that are no longer based on physical outflow velocities and photodissocation rates. The column density profiles of some species, most notably C$_2$, require multi-generation Haser models \citep{Odell1988}. Departures from the molecular distributions predicted by the spherically-symmetric Haser model become more evident with increasing telescope resolution and sensitivity. Anisotropic outflow is commonly seen in spectral line profiles observed at mm/sub-mm wavelengths \citep{cor14,gul15,rot21}. This results from diurnal modulation of the sublimation rates, jets and other heterogeneous outgassing effects related to surface/subsurface inhomogeneities and non-uniform illumination of the nucleus. Other well-known phenomena not described by the Haser model include asymmetric coma heating and photolysis rates, radial outflow acceleration \citep{biv11}, and solar wind pick-up of molecular cations. 

To provide a more physically realistic description of the outer coma, \citet{Festou1981} and \citet{Combi1980} introduced models in which fragments are emitted with an isotropic velocity kick upon their dissociation. Because these models are rooted in photochemistry, it requires that the formation pathway and reaction dynamics of the coma species are known, which is not the case for most fragment species, except for OH \citep{Feldman2004,Combi2004}. Both approaches are time-dependent, but are only valid for the collisionless part of the coma. More numerically-intensive, direct simulation Monte Carlo models are required for the most accurate determinations of molecular densities (and kinetic temperatures) across the complete range of coma scales \citep{Combi2004}. Such models can capture the complex interplay between nucleus shape, insolation, and outgassing patterns in 3 dimensions \citep{combi2020}. Less complex, 3D fluid dynamic models are also applicable for a wide range of conditions \citep{shou16}. Nevertheless, because of their simplicity and ease of parameterization, ad-hoc multi-component-type models, or semi-analytic treatments, remain a favored approach for interpretation of high-resolution ground-based and in-situ observations \citep{biv19,zhao20}.

In the outermost coma, solar wind interactions begin to dominate the gas dynamics, resulting in the formation of an ion tail containing cometary particles accelerated away in the anti-solar direction. Interpretation of coma ion observations can therefore require the use of models incorporating magnetohydrodynamics \citep{rau97}, although simpler, Monte Carlo treatments of the ion distribution can be appropriate in some cases \citep{lov98}. We refer to the chapter by \citet{Beth2023} for a more complete discussion of coma ion distributions. 

\subsection{Solar Radiation Field}\label{sec:sun}
The solar radiation field plays an important role in determining the excitation state of the atoms and molecules in the coma. Therefore, the quality of the input spectrum of the solar radiation field is a critical component when modeling cometary emissions. \rev{The tenuous nature of the coma environment permits solar radiation to penetrate deeply into the coma, where it plays an important role (e.g., sublimation of grains, solar pumping of molecules). The widespread presence of strong Fraunhofer lines across the solar spectrum, together with the typically narrow nature of cometary lines makes the effects of the Sun on the coma dependent on the comet's heliocentric velocity and the outflow velocity of the gas \fix{(known as the Swings and Greenstein effects, respectively)}. Therefore, solar pumping, fluorescence and maser emissions in comets are highly susceptible to the characterization of the fine structure of the solar spectrum and the specific velocities at the time of the observations}. 

Our knowledge of the solar spectrum has greatly improved in the past few decades due to spacecraft measurements (ATMOS; \citealt{abrams1996}, ACE; \citealt{hase2010}), and also via solar surveys performed with ground-based observatories \citep{wallace2003}. One of the biggest limitations of some of these databases is that they are not flux-calibrated, and the spectra can only be used to extract transmittance information. Although theoretical models \citep{tobiska2000,kurucz2000} have been extremely successful in calculating a flux‐calibrated solar continuum, their predictions of solar spectral features are still not optimum at high resolutions. To determine accurate solar optical/infrared templates, a general solution has been to combine theoretical and empirical solar databases \citep{bromley2021, villanueva2011, fiorenza2005}. At short wavelengths, the far ultraviolet (FUV) solar model by \citet{Fontenla2014} can be used to model the typical hard energetic radiation ($\lambda<$ 170~nm), which can be complemented by SOLSTICE measurements of the Sun during solar minimum in the 170--200~nm range \citep{Rottman1993} \footnote{\url{https://lasp.colorado.edu/home/sorce/data/ssi-data}}.

The solar spectrum varies significantly at shorter wavelengths (e.g., FUV, X-rays; \citealt{Huebner2015}). In particular for ultraviolet observations ($\lambda<$ 200~nm) of atomic features, it is important to scale the reference spectrum of the Sun to the exact conditions during the observations. An approach to capture the temporal evolution of the UV radiation field is to scale the solar spectrum by the daily averaged ﬂux as measured with the TIMED–SEE instrument \citep{woods1998,woods2000} for a given day of observations at 68~nm by the SUMER-averaged relative spectrum to estimate a high-resolution UV solar spectrum\footnote{\url{https://www.swpc.noaa.gov/}}. \rev{In addition, 
the daily 10.7~cm solar fluxes can be useful to estimate molecular lifetimes \citep{crovisier_OH1989}.}

\subsection{\rev{Energy Level Populations of Coma Molecules}}

The wavelengths and mechanisms for absorption and emission of radiation in cometary atmospheres are determined by the energy level populations of the coma gases. \final {According to quantum mechanics, the internal energy of a molecule can be separated into rotational, vibrational, and electronic modes of excitation \citep{Herzberg71}. Each of these modes is divided into a set of discrete energy levels, numbered $i$ in order of ascending energy.}

\rev{For a given molecule in local thermodynamic equilibrium (LTE) at temperature $T$, the energy level populations ($P_i$ --- the fraction of the total number of molecules in each level) follow a Boltzmann distribution: $P_i=g_ie^{-\frac{E_i}{kT}}/Z(T)$, where $E_i$ are the energies of the levels (with respect to the ground state), $g_i$ are their statistical weights, $k$ is Boltzmann's constant and $Z(T)=\sum_{i}g_ie^{-\frac{E_i}{kT}}$ is the partition function. In LTE, the excitation temperature T \fix{is equivalent} to the kinetic temperature $T_{kin}$, that relates to the collisional and velocity/thermal spectral profile of the molecule.}

Even in the general non-LTE case often encountered in the mid- or outer coma, analyses of molecular spectra (which, in reality, are likely to be based on a subset of the complete set of molecular energy levels) can be facilitated by considering the Boltzmann-equivalent temperature for that subset of levels ($i'$). Within a given vibronic level, the set of rotational level populations $P_{i'}$ can often be described using a single `rotational temperature', such that $T=T_{rot}$. Similarly, a `vibrational temperature' $T_{vib}$ can be used to characterize a (quasi-Boltzmann) distribution of vibrational level populations.

\subsection{The Equation of Radiative Transfer}\label{sec:transfer}

The density in a typical cometary atmosphere spans an extremely broad range (from less than 1~cm$^{-3}$ to more than $10^{22}$~cm$^{-3}$) \citep{Tenishev2008}, and the coma is exposed to strong and variable fluxes of solar radiation and particles. Therefore, the distribution of energy levels depends on a detailed balance of collisional and radiative excitation (and de-excitation) processes \citep{cro83a, weaver1984, cro87}. Departure of the level populations from LTE can occur in rotational, vibrational, or even electronic excitation modes, and will depend on many factors, including (1) the level of cometary activity (i.e., the density of collisional partners such as water and electrons), (2) the distance from the nucleus (since the density decreases exponentially with radius, and particle/radiation fluxes also vary with position), (3) the particular quantum characteristics of the species in question (including Einstein $A$ and $B$ coefficients; collisional cross sections), and (4) the wavelength of the observations, which often depends on the excitation energy of the system and, therefore, the source of excitation (e.g., thermal/non-thermal particle collisions, fluorescent pumping, energy balance of chemical reactions and solar wind particles). 

The total radiation field in a cometary coma is established by a comprehensive balance of emission and absorption processes (Figure~\ref{fig:processes}), which is summarized in the equation of radiative transfer \citep{liou2002}. Specifically, the intensity of radiation ($I_\nu$) propagating through the cometary coma is calculated by integrating the emission and absorption of radiation as a function of frequency ($\nu$):
\begin{equation}
\label{eq:radtran}
\frac{dI_\nu}{ds} = \gamma_\nu - \alpha_\nu I_\nu,
\end{equation}
where $\gamma_\nu$ and $\alpha_\nu$ are the emission and absorption coefficients of the gas, respectively, and $s$ is distance through the coma. These coefficients are derived from the Einstein coefficients of the gas in question ($A_{ij}$, $B_{ij}$ and $B_{ji}$, for a transition between the upper energy level $i$ and lower level $j$), as follows:
\begin{equation}
\gamma_\nu = \frac{h\nu}{4\pi}NP_iA_{ij}\phi_\nu
\end{equation}
\begin{equation}
\alpha_\nu = \frac{Nh\nu}{4\pi}(P_jB_{ji}-P_iB_{ij})\phi_\nu
\end{equation}
where $N$ is the number of gas particles per unit volume, $P_i$, $P_j$ are the relative populations in levels $i$ and $j$, respectively, and $\phi_\nu$ is a (normalized) function representing the frequency dispersion of the spectral line of interest (typically a Gaussian for individual, thermally-broadened lines). The $B_{ij}$ and $B_{ji}$ coefficients can be derived from the $A_{ij}$ coefficients of spontaneous emission and/or from the line intensities \citep{simeckova}. In order to relate an observed spectral line intensity (from energy level $i$) to a column density, \fix{the relative population of level $i$, $P_i$, needs to be known}. This can be obtained as a function of time ($t$) in the outflowing coma gas by solving the following differential equation:
\begin{equation}
\begin{split}
	\frac{dP_i}{dt} = -P_i \left[\sum_{j<i} A_{ij} + \sum_{j\neq i} (B_{ij}J_\nu + C_{ij}n_{gas})\right]\\
	+ \sum_{k>i}P_k A_{ki} + \sum_{k \neq i}P_k(B_{ki}J_\nu + C_{ki}n_{gas}).
\end{split}
\label{eq:excite}
\end{equation}

In this equation, $C_{ij}$ are the \rev{temperature-dependent} rate coefficients for transitions between levels $i$ and $j$ due to collisions between the gas particles \rev{(see Section \ref{sec:col_rates})}, and $n_{gas}$ is the number density of the colliding gas \rev{(typically H$_2$O). The $C_{ij}n_{gas}$ terms can be replaced by a sum over all significant colliding gases, if necessary}. $J_\nu$ is the local total radiation field, which includes contributions from gas emission, solar radiation, and thermal emission from coma dust particles and the nucleus (the latter two contributions are usually small enough to be neglected). Atoms and molecules can be produced (arising spontaneously in the coma, for example, as a result of dissociative  excitation by photons or electrons) in an energized (non-thermally excited) state. Such processes can be accounted for, \rev{if needed}, by adding a further source term to Equation \ref{eq:excite}.

A given astronomical observation of a comet, restricted to a particular wavelength regime, is usually sensitive to emission from a small subset of energy levels. In practice, it is therefore typical to solve for a restricted subset of energy level populations $P_{i'}$ of the gas in question, thus simplifying the calculation. For instance, in radio/millimeter-wave spectroscopy in a near-thermal regime, only the rotational levels of interest \fix{for a molecule can be considered}. For the case of some atomic emission lines, such as hydrogen and oxygen fluorescence in ultraviolet wavelengths (\citealt{Feldman2004,Noonan2021}, only a limited number of electronic states needs to be considered, although in more complex systems (such as nickel and iron), metastable states require a more complete treatment of the energy level structure \citep{bromley2021}. 

In broadband photometry and low-resolution spectroscopy of rovibronic emission at optical and infrared wavelengths, the rotational structure is unresolved. The rotational populations in the the ground-state do impact the total band pumping rates, since the solar spectrum at each rovibronic will be different - this is particularly notable for CN and OH \citep{cro83a}.  Simplified hybrid schemes are  commonly adopted, for example, in the case of a non-thermal distribution of rotational levels modified by pumping through a few of the most populated vibrational levels. For infrared spectroscopy, a full treatment of the relevant set of vibrational levels is required (pumped by solar radiation), while the ground rotational manifold remains `frozen' as a Boltzmann distribution \citep{cro83a}. As we will discuss further in Section~\ref{sec:fluoresce}, such simplifications facilitate efficient modeling and interpretation of cometary spectra. Care is needed, however, when deciding which levels can be safely excluded (or approximated) in the calculation of $P_{i'}$. For example, given the low coma kinetic temperatures (typically less than $\sim100$ K), one might erroneously consider only a few rotational levels of the ground vibronic state. Great complexity arises, however, as a result of the 5777~K solar radiation pump, which \fix{can excite} highly energetic vibrational modes of \fix{a} molecule. Other important sources of molecular excitation include photodissociative excitation (Section~\ref{sec:prompt}), impacts with hot electrons (Section~\ref{sec:elec_impact}), and solar wind charge exchange (Section~\ref{sec:charge_ex}), resulting in the population of highly-excited energy levels and giving rise to non-thermal emission at a variety of wavelengths.

\subsection{Solving for the Coma Radiation Field}

Non-thermal (disequilibrium) excitation occurs even close to the surface of the comet, where densities are high and thermalizing collisions are frequent. Evidence for this is observed at infrared (IR) wavelengths, where fluorescent emission from solar-pumped rovibrational levels is so strong and prevalent across the whole coma that it becomes a prime physical and chemical diagnostic. As a result of large Einstein $A$ values in the IR, vibrational fluorescent transitions occur rapidly, so despite traversing a manifold of vibrational states ($v$), a molecule spends most of its time in the ground state ($v=0$), where the rotational levels are thermalized by collisions. 

In addition to the various excitation mechanisms already mentioned, the energy level population at a given point in the coma depends on the local radiation field, $J_\nu$, which is calculated by summing the incident radiant energy received at that point from all solid angles. Since $J_\nu$ depends on the integrated emission over the entire spatial domain, Equation~\ref{eq:excite} should formally be solved iteratively until convergence of the level populations is achieved, which can be computationally demanding. When optical thickness is important, the physical conditions of the coma need to be computed radially along different trajectories (radial paths through the coma, starting at the nucleus) for multiple gas parcels. In this way, coma asymmetries can also be properly taken into account.

In the less dense parts of the coma where collisions are less frequent, non-LTE effects become increasingly important across all (rovibronic) levels, but the optical depth for photons leaving the coma is usually low (\emph{i.e.}, $\tau_\nu\lesssim1$). In that case, the stimulated emission and absorption terms ($B_{ij}J_\nu$ and $B_{ji}J_\nu$) tend to be small for most gases in their vibronic ground states, and can often be neglected. On the other hand, for some transitions of abundant atoms and molecules, $\tau_\nu$ can still be large in the non-LTE zone, so that photon trapping effects can have a significant impact on the energy level populations. In this regime, to avoid the computational burden of iteratively solving for $J_\nu$, the `escape probability' approximation is commonly used \citep[Sobolev's method; see][]{boc87} for interpretation of radio/submillimeter pure rotational observations. 

The solution to Equation~\ref{eq:excite} requires a time-dependent integration of the coupled set of differential equations describing how the population of each energy level changes as a parcel of gas moves outwards through the coma. Once the energy level populations are known as a function of radial distance to the nucleus, they can be mapped into three dimensions and ray-traced as a function of frequency using Equation~\ref{eq:radtran} to produce a model coma (spectral) image for comparison with observations. In practice, this is solved differently for each process/wavelength, taking into account the prevailing mechanism of excitation (Figures ~\ref{fig:spectral_overview} and ~\ref{fig:processes}). However, solving for solar fluorescence equilibrium for some molecules would require accounting for disequilibrium in millions of transitions, which would be an unsolvable system of millions of differential equations. For those cases, one can only solve for a single pump process and determine the cascade products (with the assumption that the ground rovibrational state is thermalized), which is the prevailing method to compute IR molecular fluorescence efficiencies for many species \citep{cro83a, villanueva2012}.

Similarly, when solving the rotational level populations, one can mostly concentrate on the ground state and only a few of the most relevant vibrational levels, permitting realistic treatments of the 3D opacities, pumps and other excitation processes. A further simplification to the (time-dependent) solution of Equation \ref{eq:excite} is to set $dP_x/dt=0$, invoking the steady-state approximation and solving for the energy level populations at each individual coma position (in up to three dimensions). The resulting equations of statistical equilibrium can be efficiently solved through matrix inversion methods, and Monte Carlo photon propagation can be employed for a physically accurate, self-consistent calculation of the coma radiation field $J_{\nu}$ \citep{van07,zak07}. The steady-state approximation is only strictly applicable when the radiative and collisional excitation timescales are much shorter than the dynamical timescale of the outflowing gas \citep{cor22}. However, it has the benefit of allowing arbitrary geometries in three dimensions, including the possibility of rapidly varying or even discontinuous physical parameters in the radial dimension.

\section{\textbf{Processes and their Diagnostics}}
\label{sec:processes}

\begin{figure}[t]
\centering
\includegraphics[width=\columnwidth]{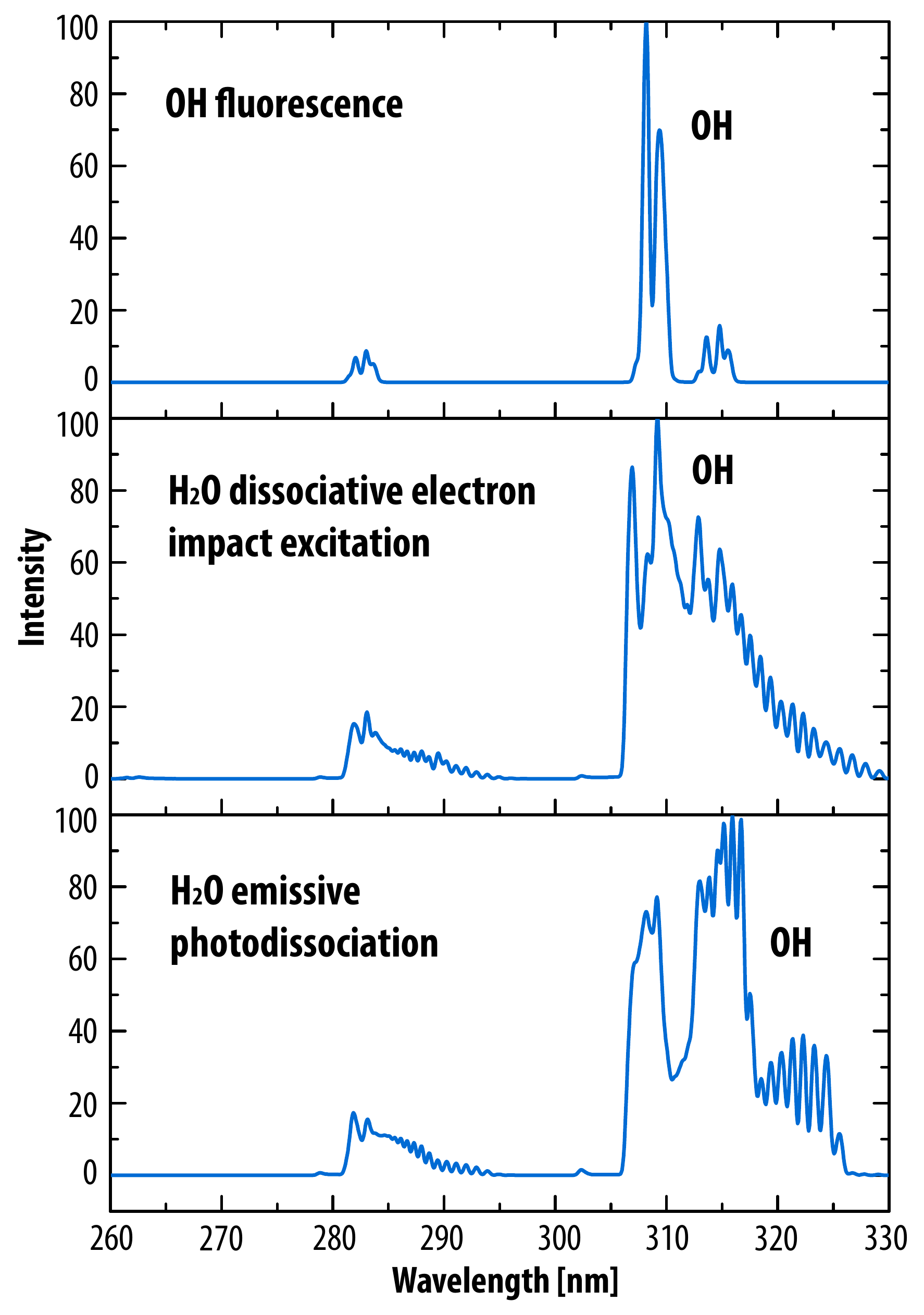}
\caption{Emission from transitions between the first excited electronic band of hydroxyl (OH) and its ground state is a prominent feature of cometary spectra in the Near-UV. Different excitation processes lead to distinct spectra. The top panel shows the fluorescence spectrum of OH radicals already present in the coma, whereas the middle and bottom figure show the emission resulting from the production of excited OH by the dissociation of water molecules by electrons and photons, respectively. Adapted from \citet{Bodewits2019}. \label{fig:OH}}
\end{figure}

The coma spans a broad range of excitation regimes and is subject to strong spatial variations in the radiation and collisional conditions. In the inner regions of the coma, collisions with neutrals and electrons provide a thermalizing influence. As we approach the more tenuous outer regions, or regions with strong chemical and/or solar pumping, additional, non-thermal excitation terms are required. For the example case of H$_2$O, some of the main processes leading to observable emission from cometary comae are summarized in Table~\ref{tab:process}. These different processes can lead to distinctly different spectra for the same molecule, as is illustrated for the OH $A~^{2}\Sigma^{+}$-- $X~^{2}\Pi$ rovibronic emission in the near-ultraviolet (Figure~\ref{fig:OH}). Fluorescence of OH radicals leads to relatively narrow emission features from the (0-0), (1-0), and (1-1) vibrational bands \citep{Schleicher1982}, whereas the direct production of excited OH by the dissociation of water molecules leads to the population of higher vibrational states. 

\begin{table*}
\begin{center}
\caption{Summary of the main processes leading to emission from water molecules and its fragments.\\}
\begin{tabular}{ l r c l }
\hline
 Collisional excitation: & ${\rm H_2O} + {\rm H_2O}$ &$\rightarrow$& ${\rm 2H_2O}^*$ \\ 
 Radiative pumping and fluorescence: & $\gamma + {\rm H_2O}$ &$\rightarrow$&  ${\rm H_2O}^*$ \\ 
 Photodissociative excitation: & $\gamma + {\rm H_2O}$&$ \rightarrow $&${\rm OH}^* + {\rm H}; {\rm O}^* + {\rm H}_2$\\
 Dissociative electron impact excitation:&${\rm e}^- + {\rm H_2O}$&$\rightarrow$&${\rm OH^*} + {\rm H}\ +$\ \ ${\rm e}^-$\\ 
 Charge exchange: & ${\rm O}^{7+} + {\rm H_2O}$&$ \rightarrow$& ${\rm O}^{6+, *} + {\rm H_2O^+}$\\
\hline
\end{tabular}

\label{tab:process}
\end{center}
\end{table*}

In the following sections, we describe the dominant excitation and radiative processes in detail for the general case of coma molecules, atoms and ions, with an emphasis on recent developments in theory, analysis methods and observational data.






\subsection{\rev{Thermal (Collisional) Excitation}}\label{sec:rot_excit}

The distribution of energy level populations for coma gases is governed by an intricate balance of radiative and collisional processes, the understanding of which starts with two basic assumptions. First, molecules sublimate with vibrational and rotational temperatures equal to the sublimation temperature of the gas source (nucleus, icy grains, or possibly larger icy chunks ejected into the coma), i.e. $\sim$ 150--200 K at a heliocentric distance $R_{h}\sim1$~au.\rev{However, collision rates are generally too small to maintain the initial vibrational population, so this population decays radiatively to the ground vibrational state}. Any further vibrational and/or electronic excitation usually occurs via non-thermal processes. Second, in the inner (often referred to as `collisional') coma, densities decrease with distance from the nucleus due to near-adiabatic expansion, but remain high enough for collisions to efficiently excite and thermalize the rotational energy level populations in the ground vibrational state. As a result, a Boltzmann distribution can be maintained with a rotational temperature similar to the kinetic temperature of the gas. Pure rotational transitions from the thermally excited rotational levels give rise to emission lines at mm/sub-mm wavelengths. In some cases, these rotational transitions may be optically thick, which reduces the efficiency of radiative de-excitation, thereby helping maintain the Boltzmann populations.

\subsubsection{Collisional Excitation and De-excitation Rates}\label{sec:col_rates}

\rev{Interpreting rotational spectra and understanding the intricacies of collisional excitation among various coma species} requires knowledge of the state-to-state collisional excitation/de-excitation rate coefficients ($C_{ij}$), which are generally less well known (and more difficult to derive) than the radiative decay coefficients. Thanks to studies of rotational excitation in interstellar clouds, collision rates between polyatomic molecules and atomic/molecular hydrogen/helium are now known for many species \citep{rou13,van20}. In contrast, comparatively little data exists for collisions of known coma molecules with the most abundant cometary gases (H$_2$O, CO, and CO$_2$). Experimental collision rates are lacking due to the difficulty of quantum-state-resolved laboratory cross section measurements. Dedicated theoretical efforts, however, are beginning to address this knowledge gap, incorporating various simplifying assumptions to allow the molecular dynamics to be solved with sufficient accuracy for application to cometary spectra.

For example, \citet{buf00} used a semiclassical method to calculate $C_{ij}$ rates for the H$_2$O--H$_2$O collisional system, which was recently revisited in more detail by \citet{bou20}. Collision rates for para-H$_2$O--HCN were calculated by \citet{dub19} using a partially-converged, coupled-states calculation, whereas for the H$_2$O--CO system, \citet{fau20} used the statistical adiabatic channel method (SACM; \citealt{lor18}). \rev{Collision rates for CO--CO (up to $J=5$) were recently published by \citet{cor22}, which are of particular use for comets at large heliocentric distances where the coma CO/H$_2$O ratio can exceed unity}. Theoretical calculations for other collision systems relevant to cometary comae are expected in the coming years, using mixed quantum/classical and SACM approaches. Presently, however, it is common to assume that the required collision rates with H$_2$O are similar to the already-known rates with H$_2$ (or He) \citep[{e.g.}][]{rot21b}, or with the additional assumption that the collision rates scale in proportion to the `reduced mass' of the colliding system \citep{hog09}. If quantum state specific $C_{ij}$ values are unavailable, an alternative approach is to assume each molecular collision has a thermalizing effect on the distribution of rotational levels, employing a nominal estimate for the gases collisional cross sections \citep{boc94,biv99,boi14}. 

For atomic species with metastable states that have long lifetimes (e.g., \ion{O}{1}, \ion{C}{1}, \ion{N}{1}, Section~\ref{sec:prompt}), it is also critical to assess the relative importance of collisional quenching and radiative decay.

\subsubsection{Rotational Temperatures:\rev{Measurements and Interpretation}}\label{sec:temp}

\begin{figure*}
\centering
\includegraphics[width=6.9in]{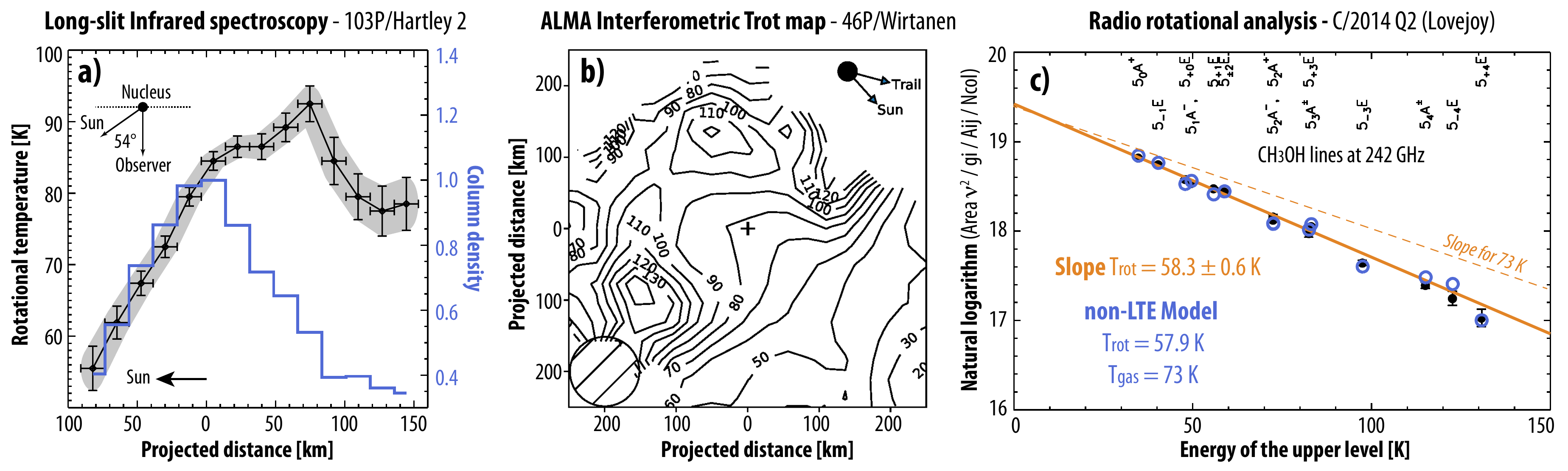}
\caption{Left: H$_2$O gas column density, ($N_{col}$, blue histogram) and rotational temperature ($T_{rot}$, black circles) distributions in 103P/Hartley 2 \citep{bonev2013}. Such spatially-resolved measurements provide a test-bed for coma thermodynamics models because the distributions of the measured parameters are diagnostic for the competition between near-adiabatic expansion cooling and various heating processes in the coma. Center: Contour map of CH$_3$OH coma rotational temperature ($T_{rot}$) in the inner coma of comet 46P/Wirtanen based on observations with the Atacama Large Millimimeter/submillimeter Array \citep{cor19b}; the coma center is denoted with a `+'. The angular resolution (hatched circle) is shown lower left. Contour labels are in units of Kelvin. Right: Rotational diagram for CH$_3$OH lines observed around 242~GHz in the coma of comet C/2014 Q2 (Lovejoy) by \citet{biv15} (black filled circles), with line of best fit (orange) denoting an inferred slope consistent with a rotational temperature of 58.3 K. Open blue circles show the calculated fluxes for individual lines based on a non-LTE radiative transfer model at a kinetic temperature of $T_{kin}=73$~K, including radiative processes and collisions with H$_2$O and electrons --- radiative cooling results in $T_{rot}<T_{kin}$. The model nicely predicts a rotational temperature matching the one derived from the slope. \label{fig:rotational}}
\end{figure*}

Gas rotational temperatures are commonly measured using radio and near-IR spectroscopy. To derive $T_{rot}$ from pure rotational lines, the individual, spectrally-integrated rotational line intensities are plotted with respect to energy of the upper state of the corresponding transition, $E_u$ \citep[{e.g.}][]{boc94}. The slope of the linear relation between these two is then directly related to the rotational temperature; see Figure~\ref{fig:rotational}. At the high spectral resolution typically available in radio, modeling the profiles of individual velocity-resolved lines in a rotational band can provide further insight into opacity and line blending effects, allowing more robust the determinations of rotational temperatures \citep{cor17}.

\citet{bonev2014} summarize methods and sources of uncertainty in the rotational temperature (usually in the ground vibrational state) from rovibrational data at IR wavelengths. Methods include (1) global fits finding the best agreement between measured and synthetic spectra (Levenberg–Marquardt $\chi^2$ minimization and correlation analysis) and (2) plotting the ratio of observed-to-modeled line fluxes on a diagram as a function of rotational excitation energy. The model $T_{rot}$ value is then varied until the points on the diagram lie along a flat line, also minimizing the line-by-line variance (zero slope excitation analysis and the ratio between flux and the g-factor $g(T_{rot})$ variance minimization). These iterative approaches for determining rotational temperatures are applicable to cases where lines from several bands are analyzed together.

In the case of high angular-resolution observations (for example with optical/IR spectroscopy or sub-mm interferometry at long baselines with the \emph{Atacama Large Millimeter/submillimeter Array}), the small ($\lesssim1''$) beam sizes (corresponding to distances $\lesssim360$~km from the nucleus at a geocentric distance of $\Delta=1$~au) are mostly sensitive to the collisionally-dominated, inner coma (Figure~\ref{fig:rotational}). Because rotational levels in the ground-vibrational state can often be characterized by a single rotational temperature, the measured $T_{rot}$ reflects the thermal state of the gas and, as such, provides a powerful diagnostic into the coma environment and processes responsible for coma heating\rev{(for example, photochemical heating, in which fast dissociation products transfer kinetic energy to the ambient coma gas; \citealt{com88})} and \rev{near-adiabatic expansion cooling}. 

We note that when several molecules are measured simultaneously across the same region of the inner coma, they all tend to share a similar rotational temperature (see for example, \citet{lippi2021}). This further demonstrates the link between $T_{kin}$ and $T_{rot}$ in the collisional coma, which, depending mainly on gas production rates, can extend from  tens to thousands of kilometers from the nucleus.

Measuring $T_{rot}$ is critical for deriving column densities from an incomplete set of molecular line observations, as is typically the case given the limited wavelength coverage and sensitivity of astronomical observations. In that case, $T_{rot}$ provides a useful approximation for the populations of unobserved levels, so that the total molecular number density can still be obtained. 

\subsubsection{\rev{Departure from LTE in the Outer Coma}}\label{sec:nonLTE}

As the coma density falls with increasing distance from the nucleus, the reduced collision rates become insufficient to maintain LTE, so the rotational level populations in the ground vibrational state start to deviate strongly from a thermal (Boltzmann) distribution. The interpretation of measured $T_{rot}$ becomes more complex, and its close relationship with $T_{kin}$ breaks down. This transition from LTE to non-LTE regimes is commonly encountered in mm/sub-mm non-interferometric (i.e., `single-dish') rotational spectroscopy, with telescope beam sizes larger than a few arcseconds (corresponding to radial distances $\gtrsim1000$~km from the nucleus, for a geocentric distance of 1~au). As a result of relatively low telluric opacities and the high (milli-Kelvin) sensitivity of detectors, such observations represent the primary ground-based technique for the detection of new, complex coma molecules \citep{biv15,Biver2023}. In contrast to near-IR spectroscopy and sub-mm interferometry at long baselines, these single dish measurements represent an average signal from the collisional and the extended coma, hence, interpretation of pure rotational lines must account for the departure from LTE of the rotational populations with increasing nucleocentric distance. In addition to a knowledge of the collisional excitation rates (Section~\ref{sec:col_rates}), understanding this departure from LTE involves the processes of radiative excitation (pumping) and fluorescence, discussed in the next two sections.

\subsection{Radiative Excitation (Pumping)}\label{sec:pumping}

\rev{Coma gases interact with various radiation fields. The most important one is the solar radiation field. Additional, minor sources of radiation include scattered and thermal emission from the nucleus and from the coma dust, emission from excited coma molecules, and the 2.7~K cosmic background radiation. This non-thermal excitation process is often referred to as radiative pumping. Absorption of solar photons is by far the major process leading to vibrational pumping (caused by IR photons) and electronic excitation (caused by UV and optical photons). Vibrational excitation from thermal dust emission should also be considered at wavelengths longer than $\sim$ 10~$\mu$m$ $~ \citep{cro83a}. Unlike vibrational and electronic excitation, pure rotational pumping by solar photons is negligible because the solar flux is too weak at the long (millimimeter/submillimeter) wavelengths of the rotational transitions. On the other hand, rotational levels (in the ground vibrational state) can be excited by the 2.7 K cosmic background \citep{boc_comets2}}.

\subsection{Fluorescence} \label{sec:fluoresce}

\rev{Radiative excitation is the first step in the fluorescence process. An atom or molecule is `pumped' to an excited level due to absorption of a solar photon (Section~\ref{sec:pumping}), followed by radiative decay via spontaneous emission. Fluorescence impacts the interpretation of all types of spectra from coma gases because: (1) This process explains the redistribution of rotational level populations in the extended (non-LTE) coma, as needed to interpret rotational spectra, especially from single-dish observations, and (2) Solar-induced fluorescence is the main process leading to emissions from virtually all comet gasses in a broad wavelength range spanning from the ultraviolet to the infrared}.

\subsubsection{Transition from Thermal to Fluorescence Equilibrium}

\begin{figure}[t]
\centering
\includegraphics[width=\columnwidth]{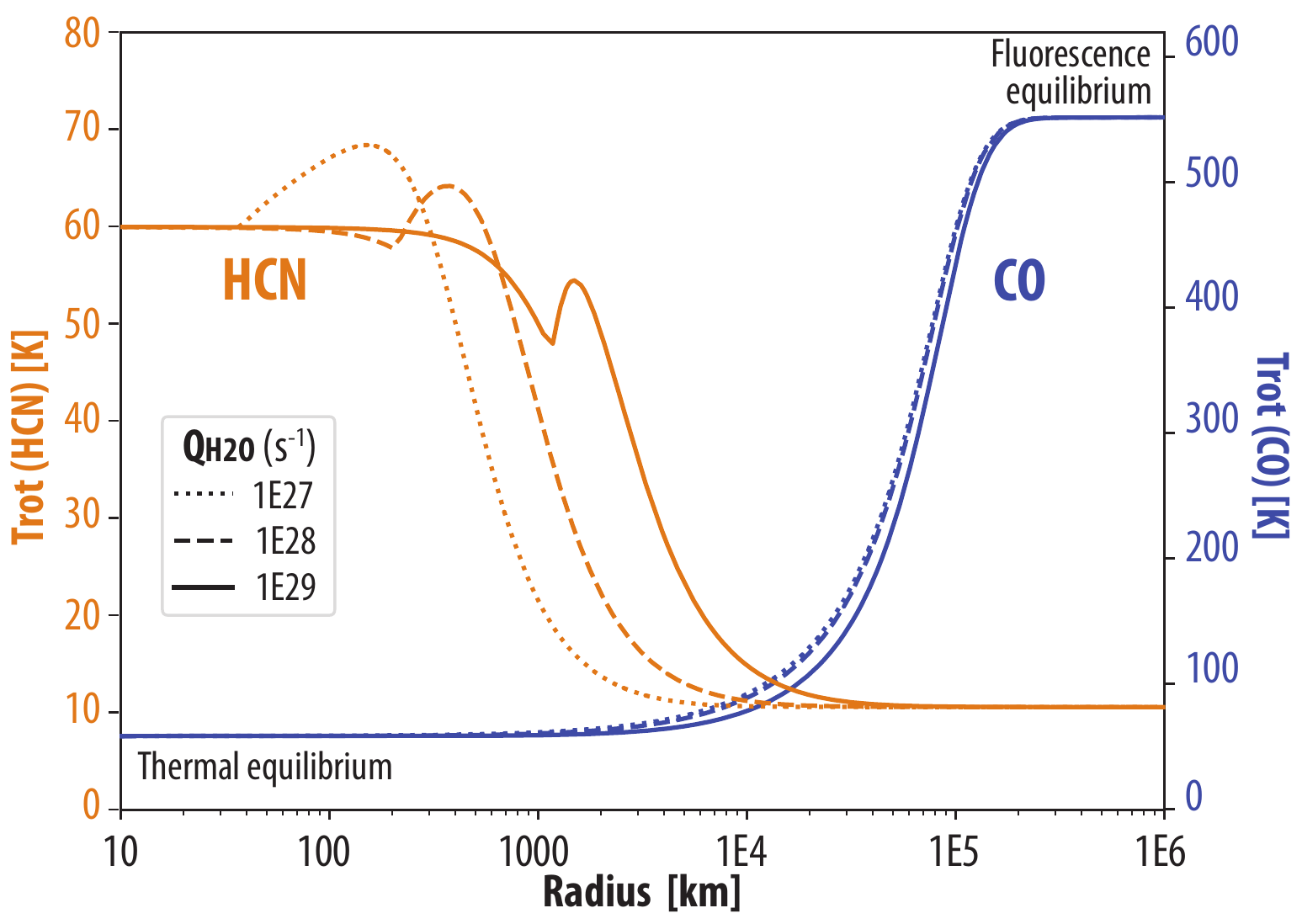}
\caption{Simulated rotational temperatures ($T_{rot}$) of HCN (orange; left ordinate) and CO (blue; right ordinate) as a function of distance from the nucleus, for three different water production rates ($Q_{\rm H_2O}$; 10$^{27}$~molecules/s --- dotted, 10$^{28}$ molecules/s --- dashed, 10$^{29}$ molecules/s --- solid), based on the populations of the lowest seven energy levels for each molecule. This assumes a spherically-symmetric outflow, with constant gas kinetic temperature $T_{kin}=60$~K, and a heliocentric distance of 1~au. Rotational energy level populations were calculated using Equation~\ref{eq:excite}, with CO--H$_2$O collision rates from \citet{fau20}, and HCN--H$_2$O collision rates from \citet{dub19}. Rotational temperatures evolve from thermal equilibrium at high density (on the left) to fluorescence equilibrium in the low-density outer coma (on the right). For additional model details, see \citet{cor22}. \label{fig:trot}}
\end{figure}

\begin{figure*}[t]
\centering
\includegraphics[width=6.5in]{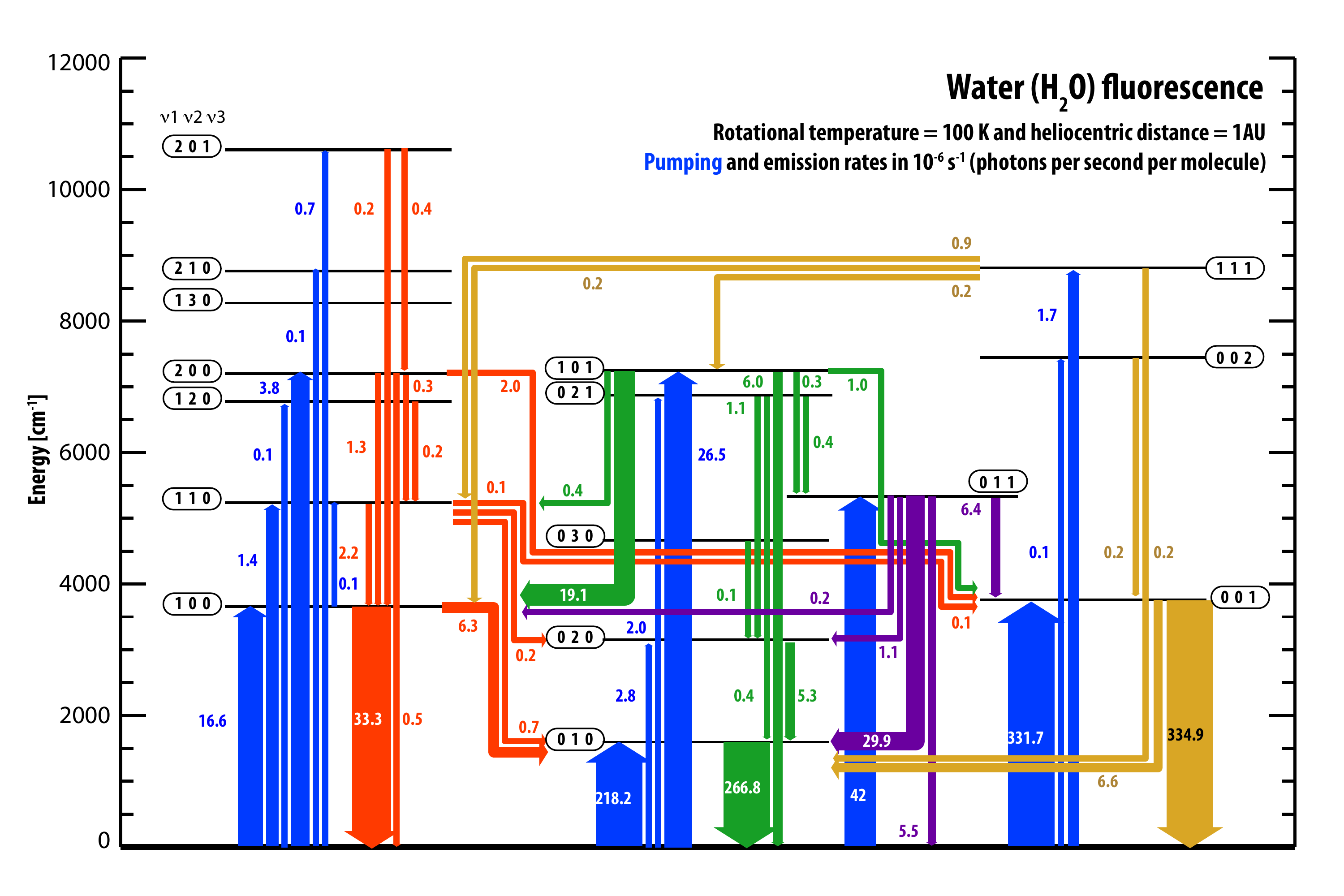}
\caption{Diagram showing the full non-resonance fluorescence tree for H$_2$O in a comet at 1~au from the Sun with a rotational population at 100~K. The pumping rates (in units of photons per second per molecule, shown in blue) were calculated considering a realistic solar model, and the emission rates (shown in red/green/purple/yellow colors) were calculated by subsequent cascade down to the ground-vibrational level and considering line-by-line and level-by-level branching ratios, which take into account all 500 million transitions. Adapted from \cite{villanueva2012}.\label{fig:metro} }
\end{figure*}

\rev{As molecules sequentially absorb solar photons and radiatively decay via spontaneous emission they undergo transitions, in which the total angular momentum ($J$) can change by an amount equal to $\Delta{J}=0$ or $\pm1$. This leads to a change in the degree of rotational excitation. For an initial distribution of rotational level populations $P_{i,v'}$ in vibrational state $v'$, following a vibrational (or electronic) transition, the $\Delta{J}$ selection rules result in characteristic $P$, $Q$ and $R$ branches in the spectrum, corresponding to $\Delta{J}=-1$, 0 and $+1$, respectively. The rovibrational transition probabilities vary as a function of $J$, $\Delta{J}$, the dipole moment, and the Franck-Condon factor of the vibrational transition \citep{Herzberg71}. Therefore, after undergoing repeated absorption and emission of photons as a gas parcel travels outward through the coma, the distribution of rotational levels $P_{i,v'}$ changes from the initial, Boltzmann distribution prevalent in the inner coma. At large distances from the nucleus a balance between absorption and spontaneous emission is ultimately realized and the distribution of rotational levels attains fluorescence equilibrium with the solar radiation field}. 

\rev{The change from LTE to non-LTE distributions is evident in observed rotational temperatures. Spatially-resolved measurements of the rotational temperature of H$_2$O acquired with the \emph{Spitzer Space Telescope} showed a decrease with distance from the nucleus, best described as a transition from thermal to fluorescence equilibrium \citep{woodward2007}. Importantly, this transition occurs more rapidly for strongly polar molecules, such as HCN, with higher rotational transition rates (larger Einstein $A$ values), whereas molecules with smaller dipole moments, such as CO, maintain LTE out to much greater distances from the nucleus. Figure~\ref{fig:trot} shows the theoretical evolution of CO and HCN rotational temperatures with distance from the nucleus for comets with increasingly large water production rates ($Q_{\rm H_2O}=10^{27}$ -- $10^{29}$ molecules\,s$^{-1}$).  The smaller Einstein $A$ values for rotational transitions of CO (for a given $J$ level) lead to a larger population of the higher $J$ states as a result of solar pumping. Consequently, at fluorescence equilibrium in the outer coma of a comet at 1~au from the Sun, $T_{rot}$(CO) reaches 550~K, whereas $T_{rot}$(HCN) reaches only 11~K. The relatively large rate of collisions between HCN and electrons is also responsible for significant departures of $T_{rot}$(HCN) from the kinetic temperature of the neutral gas, seen as humps in the $T_{rot}(r)$ curve in Figure \ref{fig:trot}, between 100 and 2000~km from the nucleus, where the densities of (hot) photoelectrons in the coma begin to become significant. The dramatically different behavior of these two commonly observed coma molecules demonstrates the importance of considering the detailed microphysics when using limited observational data for the derivation of production rates}.

\subsubsection{Advancements in Fluorescence Models}\label{sec:fluo_mod}

Fluorescence gives rise to emission lines of many cometary volatiles of significance to both cosmogony and astrobiology, including water (H$_{2}$O, see Figure~\ref{fig:metro}, and HDO), symmetric hydrocarbons (CH$_{4}$, C$_{2}$H$_{2}$, C$_{2}$H$_{4}$, C$_{2}$H$_{6}$, oxidized carbon compounds (CO, H$_{2}$CO, CH$_{3}$OH, OCS), nitrogen species (NH$_{3}$, HCN), and daughter species (e.g., C$_{2}$, CN, OH, NH).
In particular, symmetric hydrocarbons can only be detected through their fluorescence IR emission, due to the lack of a permanent dipole moment. High-resolution ($\lambda$/$\Delta\lambda$ $>$ 20,000) optical and near-IR spectrographs at ground-based facilities can detect many volatiles simultaneously via multiple lines of each species \citep{Cochran2012, boattini}.

The interpretation of observed spectra requires the development of fluorescence models. The first fluorescence calculations focused on the electronic excitation of radicals such as OH, CN, and CO$^+$ at optical and near-UV wavelengths \citep{Schleicher1982, Schleicher1983, Magnani1986}. The characterization of the IR region in comets has grown tremendously in the last decade, thanks to the recent advent of extensive ab-initio high-energy spectroscopic linelists (ExoMol, \citealt{exomol}), and more complete experimental linelists of trace species and hydrocarbons (HITRAN, \citealt{hitran}). Because fluorescence involves pumping and cascades from all rovibrational levels, the calculations require handling of a large number of high-energy transitions (due to solar pumping) as well as low-energy (thermal) transitions. This is particularly relevant for non-resonance fluorescence, which dominates the IR emissions of H$_2$O probed with ground-based observatories. Modern fluorescence models ingest information from many spectroscopic sources and handle the billions of transitions needed for accurate branching ratio calculations \citep{disanti2006, villanueva2012, villanueva2013, gibb_ch3d, kawakita2011}. For instance, a recent high-energy water database contained $5\times10^8$ transitions for H$_2$O and $7\times10^8$ for HDO \citep{villanueva2012}. Figure~\ref{fig:metro} presents the main elements in a fluorescence calculation (pumps, cascades, and branching ratios), all underpinned by a complete molecular Hamiltonian and an equilibrium non-LTE model. These new models have become more accessible to the community by the development of open and public repositories and the inclusion of non-LTE modeling and fluorescence capabilities via public online tools \citep{psg2018}. 

The interpretation of fluorescence spectra requires a hybrid approach, starting with the generally good approximation that in the inner coma rotational populations in the ground vibrational state are collisionally equilibrated to a Boltzmann distribution, and then incorporating the complexity of non-LTE physics. Solar pumping raises a molecule to a highly-excited vibrational state (Section~\ref{sec:pumping}) through rovibrational transitions, based on the impinging solar flux ($J_v$) and the Einstein $B_{ji}$ coefficients of those transitions. Then, the molecule quickly decays radiatively to intermediate vibrational levels (non-resonant fluorescence or `hot-bands') or directly back to the ground vibrational state (fundamental bands). Each cascade gives rise to observable optical/IR lines, which can be characterized by a corresponding fluorescence efficiency. The cascade process is followed sequentially for all rovibrational levels, until the molecule returns to the ground vibrational state. At each level, the direction and proportionally of the cascade is calculated based on the branching ratios, which are determined from the spontaneous emission coefficients ($A_{ij}$) for transitions from that level. Fundamental bands with direct counterparts in absorption in the Earth’s atmosphere (such as CH$_{4}$, CO, HDO) require geocentric Doppler shifts for their detection in ground-based studies of comets. H$_2$O vibrational hot-bands are the best ground-based diagnostic for water, because these transitions originate from levels not significantly populated in Earth’s atmosphere \citep{mumma1986}. The final product of IR fluorescence models is the emission efficiency (g-factor, in units of photon~s$^{-1}$ molecule$^{-1}$, or W~molecule$^{-1}$) for each rovibrational line across all possible vibrational modes.

A challenge in fluorescence calculations is the inclusion of opacity effects and non-Boltzmann rotational level populations, which require solving the radiative transfer equilibrium states iteratively. In most cases, the pumping transitions are optically thin due to the transient and sporadic nature of high-energy rovibrational transitions, yet for highly active comets, the coma can be quite opaque for the pumps of some abundant species. A hybrid solution to this issue can be achieved by scaling the line fluxes based on the probability of the corresponding pump being opaque, which can be approximated by exploring the Einstein $B_{ji}$ coefficient and average densities in the atmosphere. Integrating this approach into the $Q$-curve formalism (i.e., extracting production rates based on flux collected in successive apertures along the slit; \citealt{DR1998}) has helped in addressing optically thick pumps in the derivation of densities and rotational temperatures in comet C/2006 W3 (Christensen) \citep{bonev2017}.

\fix{\citet{bock_HB_2010} demonstrated the importance to consider optical depth effects in interpreting spatial distributions from infrared observations in exceptionally bright comets, such as C/1995 O1 (Hale-Bopp), which exhibited extraordinary high CO production rates ($\sim 10^{29} - 10^{30}$ molecules s$^{-1}$). This work presented radiative transfer calculations of CO fluorescent line intensities from the 1--0 band, taking into account (1) opacity in the solar pumps, (2) radiation trapping caused by self-absorption of infrared photons emitted by CO molecules in coma, which acts as an additional source of vibrational excitation, counterbalancing the reduced direct pumping by solar photons, and (3) attenuation of the observed infrared emission, also due to self-absorption. The authors showed that the spatial profiles of CO in Hale-Bopp can be explained by these optical depths effects rather than indicating the presence of extended coma sources of CO, as inferred by earlier analyses reviewed by \citet{boc_comets2}}.

\fix{Treating opacity effects in the coma is also particularly relevant when interpreting cometary spacecraft data, since the measured bands tend to be the strong fundamentals, the densities sampled of the inner coma are particularly large, and the effects regarding the directionality of the excitation source become notable. As such, dedicated models that combine modern numerical and radiative transfer methods have been developed to model such spectra \citep{Gersch2014,Gersch2018,Debout2016, cheng_opr22}}.

\subsubsection{Coma Diagnostics from Fluorescence Spectra}\label{sec:fluo_diagnostics}

In the infrared, comparison of measured spectra with fluorescence models enables the determination of gas rotational temperatures (Section ~\ref{sec:temp}) and column densities ($N_{col}$), which in turn can be a test-bed for coma thermodynamic models. \citet{fougere2012} compared synthetic $T_{rot}$ and $N_{col}$ spatial profiles predicted by a kinetic Direct Simulation Monte Carlo model \citep{Tenishev2008} with the long-slit spatial distributions of these parameters obtained from IR fluorescence spectra in 73P/Schwassmann-Wachmann~3B. Their work suggested that water sublimated from icy grains in a rotationally hot state imparts energy through collisions with the ambient coma, thereby offsetting the cooling due to near-adiabatic expansion. This scenario also explains the elevated $T_{rot}$ in the projected anti-sunward direction in 103P/Hartley~2 (Figure~\ref{fig:rotational}, left panel), where the EPOXI mission observed that icy grains were a significant source of water in the coma \citep{ahearn2011, protopapa2014}. 

Comparing the spatial distributions of column density among different volatiles addresses whether these species are associated with common or distinct outgassing sources \citep{Biver2023}. Major differences in spatial distributions point to heterogeneous outgassing reflecting entirely distinct sources or one or more additional source(s) for some volatiles \citep{kawakita2013, dello2022}. \citet{disanti2018} showed that short-term (hours) temporal variations in coma relative abundances can be directly linked to heterogenous outgassing, as revealed by spatial distributions of fluorescent emission measured in comet C/2013 V5 (Oukaimeden). Another notable example is C/2007 W1 (Boattini), for which \citet{boattini} interpreted the emission profiles in terms of release from two distinct moieties of ice: (1) clumps of mixed ice and dust released from the nucleus into the sunward hemisphere, and (2) small grains of nearly pure polar ice (water and methanol, without dark material or apolar volatiles). 

Improvements in fluorescence models have also resulted in more accurate measurements of spin ratios in cometary comae. An intrinsic property of molecules is the existence of distinct symmetries of nuclear spin species, which do not interact radiatively under electric-dipole selection rules. Inter-conversion among these spin species is strongly forbidden in the gas, leading to long persistence times in the coma. An often overlooked aspect is that the fluorescent g-factors have separate dependencies on rotational temperature and on spin ratio. For example, even at the same $T_{rot}$, spectra of water will look substantially different for ortho-H$_2$O to para-H$_2$O abundance ratios of 1.5 and 3.0. Without factoring the effects of rotational temperature and spin ratio into the modeled g-factors, the observed line strengths cannot be accurately reproduced, and this can influence calculated column densities and production rates.

\begin{figure*}
\centering
\includegraphics[width=6in]{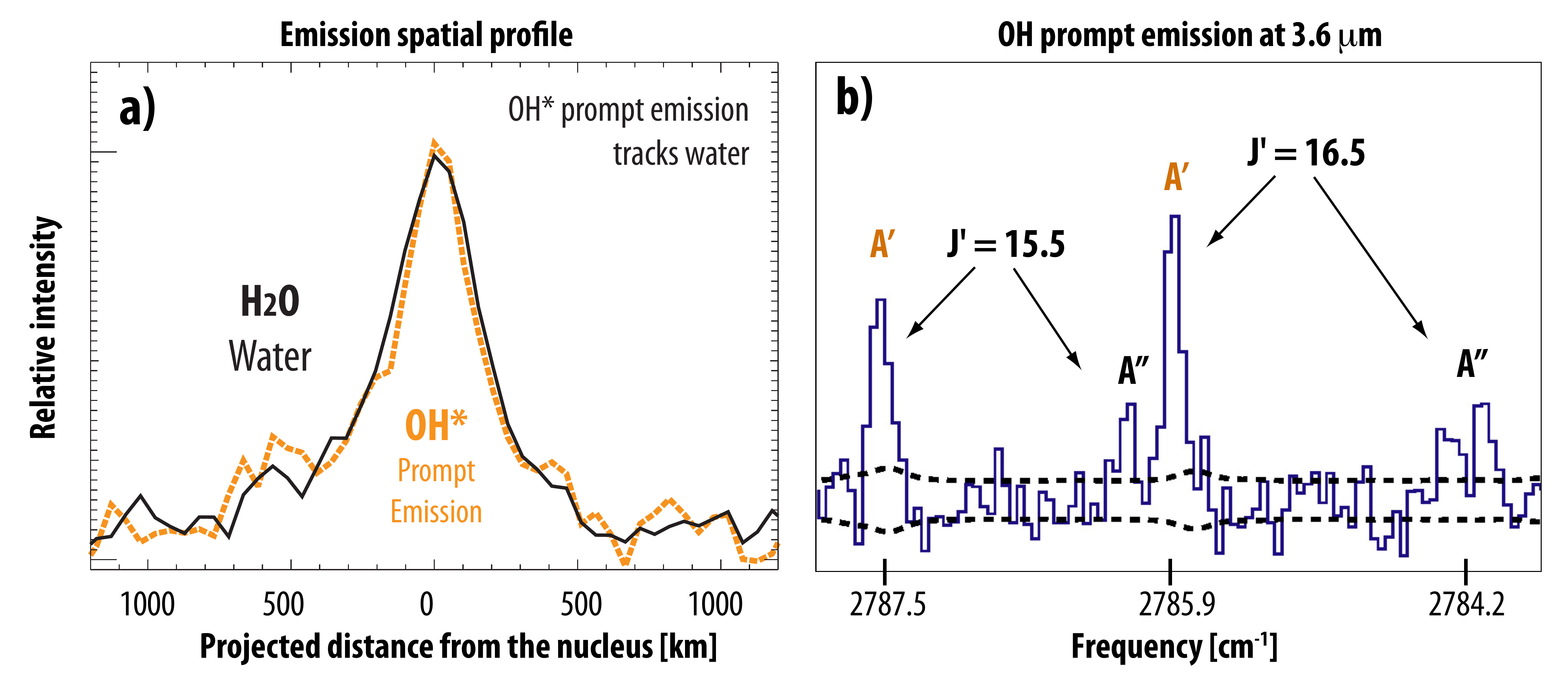}
\caption{a) Spatial distribution of H$_{2}$O and of OH prompt emission intensity in C/2000 WM1. Prompt emission commonly peaks at the nucleus and tracks the spatial distribution of the dissociative precursor. b) Example of OH prompt emission lines often detected in comets. Near-IR measurements typically use frequency in wavenumber ($cm^{-1}$) units (2785.9 $cm^{-1} \approx$ 3.6 $\mu$m). $J'$ is the rotational quantum number of the upper state for a transition. The ratio between the emission efficiencies of the $\Pi(A')$ and $\Pi(A'')$ $\Lambda$-doublet components (marked as $A'$ and $A''$ on the plot) is an important diagnostic for the H$_{2}$O dissociation dynamics leading to rovibrationally excited OH states (after \citealt{Bonev2006, BonevMumma2006}). \label{fig:prompt}}
\end{figure*}

Infrared fluorescence spectra are particularly suitable for obtaining spin ratios because of the simultaneous sampling of numerous lines from different spin species and molecules provided by modern high-resolution spectrometers. Because the lowest energy levels in the various spin ladders differ slightly, the ratio between the total populations of different spin states would be temperature-dependent if we assume that the given spin ratio was realized in thermal equilibrium. Thus, a nuclear spin temperature, $T_{spin}$, can formally be derived from measured spin ratios. Although a spin temperature of 30~K was suggested by observational studies of H$_2$O and NH$_3$ in some comets \citep{dello2005, kawakita2006, mumma2011}, subsequent ground-based retrievals of $T_{spin}$ employing the latest models typically indicate spin temperatures above 50~K, consistent with statistical equilibrium \citep{villanueva2012, bonev2013}.\fix{\citet{cheng_opr22} examined the effects of optical depth in retrieved spin ratios from water fluorescent emission detected by the Visible and Infrared Thermal Imaging Spectrometer (VIRTIS-H) on board Rosetta, also concluding that statistical equilibrium $T_{spin}$ best explains their measurements.}

Currently, there is a strong debate over whether $T_{{spin}}$ preserves a record of the formation temperatures of cometary volatiles. Recent laboratory measurements do not support this scenario and explore the possibility that phase transition phenomena \citep{hama2016}, energetic particle irradiation \citep{sliter2011}, and/or formation of clusters \citep{tanner2013} may reset the ortho/para ratio of cometary water. The situation for ammonia is not well characterized experimentally, while the exchange of H-atoms in CH$_4$ and other hydrocarbons (C$_2$H$_6$, CH$_3$OH) should be extremely slow even in ices.

\subsection{Photodissociative Excitation}
\label{sec:prompt}

In this non-LTE process, parent species (H$_2$O, H$_2$CO, CO$_2$, CO) are excited into unstable dissociative states by solar UV photons, producing photodissociation fragments (OH, CO, H$_2$, \ion{O}{1}) in excited states. For example, photolysis in the second absorption band of H$_2$O (primarily by Ly-$\alpha$ radiation) can lead to OH in electronically excited states ($A\ ^{2}\Sigma^{+}$) or in vibrationally and rotationally excited levels within the ground electronic state ($X\ ^{2}\Pi$), while photolysis in the first absorption band ($\lambda >$ 136 nm) leads predominantly to rovibrationally excited OH ($X\ ^{2}\Pi$) \citep{crovisier_OH1989}. These states of OH cannot be efficiently populated by fluorescence or collisions and generally have short lifetimes ($\sim$ 10~ms). As a result, the dissociatively excited fragment decays radiatively giving rise to prompt emission.

This emission provides two main diagnostics. (1) Because of the short radiative lifetimes of dissociation products, optically-thin prompt emission is a good tracer for the spatial distribution of the precursor molecule (see Figure~\ref{fig:prompt}), and can also be used to approximate the production rate of the precursor \citep{Bertaux1986}. However, analysis of prompt emission requires evaluating optical depth effects in the solar UV for exceptionally active comets \citep{Bonev2006}. (2) The relative intensities of prompt emission lines may be quite different from those expected from fluorescence (Figure~\ref{fig:OH}) and can help understand the dissociation dynamics in cometary comae \citep{BonevMumma2006, Bodewits2016}. The quantum state distributions of dissociation products, inferred from comet observations, can be compared with those from laboratory studies \citep{Feldman2009, AHearn2015}. These distributions are governed by the exit channels from the dissociative parent state, whose excitation in turn depends on the initial state population of the parent (see Section~\ref{sec:rot_excit}) and especially on the energy of the UV photon (e.g., first or second absorption band of H$_2$O). 

\citet{mumma82} and \citet{Bertaux1986} predicted that prompt emission would be an important component in respectively the infrared and ultraviolet spectra of comets. \citet{crovisier_OH1989} modeled the integrated prompt emission rates for several OH vibrational bands. \citet{bock_crov89} showed that OH prompt emission will contribute to low resolution ($\lambda$/$\Delta\lambda$ $\sim$ 160) infrared cometary spectra near 2.8 $\mu$m. Early detections of prompt emission were reported by \citet{budzien1991}, \citet{brooke96}, \citet{Mumma2001}, and references therein. Here we complement the recent summary by \citet{Thomas2020}, highlighting several post-Comets~II developments. 

\citet{AHearn2015} reported the first spectrally resolved detection of OH prompt emission in the near-UV 0--0 band of the $A~^{2}\Sigma^{+}$-- $X~^{2}\Pi$ electronic transitions. They analyzed spectroscopic observations of comet C/1996 B2 (Hyakutake), whose small geocentric distance (0.11~au) allowed probing the near-nucleus region and to detection of prompt emission from the ground. The `rotationally-hot' quantum state distribution of OH measured in Hyakutake was in excellent agreement with laboratory studies of the second absorption band of H$_2$O. 

\citet{LaForgia2017} observed the near-UV OH prompt emission in narrow-band images taken during the \emph{Deep Impact} spacecraft flyby of comet 103P/Hartley~2. Demonstrating that the prompt emission closely tracked the maps of water vapor in the inner coma, the authors proposed a dedicated OH prompt emission filter, which could be used to directly image the distribution of H$_2$O in the inner coma when the first 200~km are resolved. Centered at $\sim$ 318.8~nm, such a filter would be suitable for wavelengths commonly accessible by CCD's on both ground-based telescopes and platforms in space. 

\begin{figure*}
\centering
\includegraphics[width=6in]{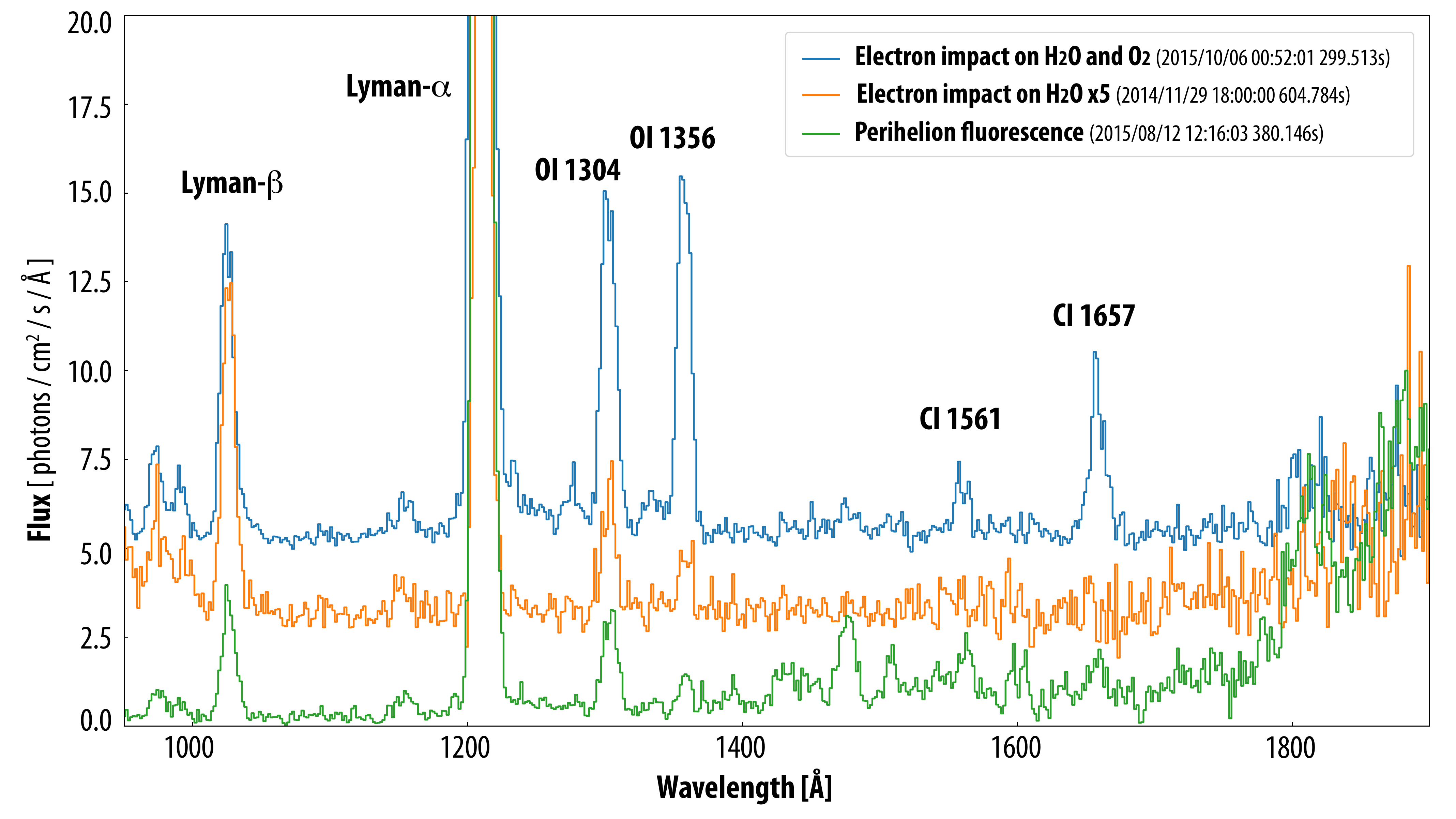}
\caption{ Three spectra from the \emph{Rosetta/Alice} ultraviolet spectrometer are shown to illustrate three unique signatures observed in the near-nucleus coma around 67P/Churyumov-Gerasimenko. All spectra are created using the narrow middle section of the Alice slit. Electron impact on molecular oxygen and water (blue) produces the clearest signature of electron impact at the \ion{O}{1}] 1356 Å emission feature. Electron impact on water (orange) is clearest in the early days of the mission when 67P was beyond 2.7~au. It has a much weaker signal in that spectral region. For comparison, a spectrum taken at perihelion, in green, shows a richer spectrum of CO Fourth Positive group features and resonance fluorescence of the Lyman series, but little in the way of \ion{O}{1}] 1356~Å (adapted from \citealt{Feldman2018}). \label{fig:electron}}
\end{figure*}

Many P-branch lines from vibrational prompt emission bands in OH ($X^{2}\ \Pi$) are detectable in ground-based near-IR observations of even moderately bright comets. Using simultaneous measurements of H$_2$O and OH, \citet{Bonev2006} empirically calibrated emission efficiencies (OH photons s$^{-1}$ per H$_2$O molecule) for a suite of OH lines (2.8 -- 3.6 $\mu$m). This has helped isolate the contribution of OH in spectrally crowded regions throughout the L-band \fix{in near-infrared studies of cometary volatiles} \citep[see][]{Biver2023}. 

The rotational distribution of OH ($X\ ^{2}\Pi$) cannot be approximated by a single Boltzmann distribution. The dissociation of H$_2$O can lead to $\Lambda$-doublet states of OH with the same rotational quantum number $J'$, but with a different symmetry of the electronic wave function: $\Pi(A')$ and $\Pi(A'')$, as defined by \citet{Alexander1988}. These states generally have vastly different populations, which affects the relative intensities of prompt emission lines (Figure \ref{fig:prompt}b). \citet{BonevMumma2006} showed that the ratio of OH line intensities from transitions originating from $\Pi(A'')$ and $\Pi(A')$ levels has a strong dependence on $J'$. This ratio may be diagnostic for the particular dissociation channel, reflecting preferential population of $\Pi(A'')$ states at low-$J'$ (dissociation in the first absorption band), and for $\Pi(A')$ states for $J' >$ 9.5 (second absorption band).

The $a\ ^{3}\Pi$ states of CO can be produced directly by the dissociation of CO$_2$ and the ensuing prompt emission has long been considered one of the main sources of the CO Cameron bands in UV spectra of comets \citep{weaver94}. However, \citet{RaghBhard2012} evaluate the importance of both photodissociatve excitation and electron impact excitation (see Section \ref{sec:elec_impact}) in populating CO ($a\ ^{3}\Pi$) and they concluded the latter is the main production mechanism in 1P/Halley.

\citet{Kalogerakis2012} showed that CO~($a\ ^{3}\Pi$) can be efficiently populated via cascades from higher-energy levels of CO. These authors identified visible and near-IR prompt emission in laboratory spectra of CO resulting from CO$_2$ photodissociation. CO was produced in $a'\ ^{3}\Sigma^{+}$, $d\ ^{3}\Delta$, and $e\ ^{3}\Sigma^{-}$ states. Cascades from these states can populate $a\ ^{3}\Pi$ and lead to prompt emission in the visible and near-infrared (wavelengths above 500 nm). \citet{Kalogerakis2012} suggested that if found in comets, these emissions would be an indirect "marker" for the presence on CO$_2$ in the coma, potentially leading to upper limits of the CO$_2$ abundances.

Dissociative excitation can lead to the production of radicals in states with longer radiative lifetimes (metastable states), which therefore do not favor prompt emission. These products can then be further excited by fluorescence. In the FUV, \citet{Liu2007} identified many lines pumped by solar Ly-$\alpha$ from rovibrational levels of H$_{2}$ produced by photodissociation of H$_2$O. \citet{Feldman2009} and \citet{Feldman2015} discussed CO and H$_{2}$ fluorescence respectively, pumped from dissociatively excited states of H$_2$CO. 

Several observational and modeling studies have focused on better understanding dissociative excitation leading to metastable atomic states. \citet{mckay2015}, \citet{Decock2015}, \citet{Raghuram2019}, and references therein discuss both the usefulness and the uncertainties in an indirect method to obtain CO$_2$ production rates based on the intensity ratio of the green ($^1$S - $^1$D; $\lambda$ = 557.7~nm) and red doublet ($^1$D - $^3$P, $\lambda$ = 630.0 and 636.4~nm) lines of O I. This is a potentially important application, because CO$_2$ is a major volatile in comet nuclei that cannot be detected from the ground, unlike the easily detectable lines of \ion{O}{1}. Two key challenges for this methodology are that atomic oxygen can be produced by the dissociation of various parents (such as H$_2$O, CO$_2$, O$_2$, and CO), and that the relevant dissociative excitation rates are not well known. These rates are difficult to constrain in the laboratory owing to the long radiative lifetime of metastable atomic states.\fix{Nevertheless, based on laboratory studies of H$_{2}$O photodissociation leading to O~$^1$S and vibrationally-excitated H$_{2}$, \citet{kawakita2022} derived H$_{2}$O photodissociation rates leading to both O~$^1$S and O\ $^1$D.} 

\citet{Raghuram2019} utilized the H$_{2}$O-depleted coma of C/2016 R2 (PanSTARSS) as a natural laboratory to study dissociative excitation of CO$_2$, CO, and N$_2$, leading to \ion{O}{1}, \ion{C}{1}, and \ion{N}{1}, respectively. Collisional quenching is being evaluated in interpreting the line ratios and line widths of forbidden transitions, measured for different projected distances from the nucleus. These ratios help constrain dissociation yields and also identify photolysis of O$_2$ as an additional source of the \ion{O}{1} red and green lines in comets. The line ratio for the \ion{N}{1} doublet ($\lambda$ = 519.8 and 520.0~nm, respectively) measured in C/2016 R2 is useful for obtaining the intrinsic transition probability ratio of the two sub-levels of N~$^2$D. Further investigations are needed to more fully investigate these complex processes.

\subsection{Electron Impact Excitation}
\label{sec:elec_impact}

As solar radiation impinges on the coma, photoelectrons arise in the gas with a distribution of energies that peaks between 1 and 30~eV \citep{Cravens1990,Engelhardt2018}. These electrons redistribute their energy via collisions with atoms and molecules \citep{Cravens1986}, and in doing so, contribute towards various coma excitation and emission processes, the nature of which depends strongly on electron energy.

Collisions with thermalized electrons in the inner coma help maintain LTE, but outside the contact surface, where the degree of ionization increases and electron temperatures and densities rise steeply with radius \citep{kor87,ebe95}, impacts with hot ($\sim10^4$~K, or a few eV) electrons can strongly affect the excitation of neutral coma gases \citep{xie92,lov04}. Similar to the case of neutral-neutral molecular collisions (Section~\ref{sec:col_rates}), the state-to-state collision rates involving electrons and neutral molecules can be obtained from detailed quantum mechanical calculations \citep{fau07}. However, such calculations are yet to be performed for most molecules of interest over the range of electron energies found in cometary comae. Consequently, the simplified formula of \citet{iti72} (using the Born approximation) is frequently used to determine state-to-state electron impact rates \citep{biv97,zak07,cor19}.

\rev{\subsection{Dissociative Electron Impact Excitation}}
 Electrons with energies above 10~eV that collide with neutral molecules in the coma can produce excited fragments. \emph{Rosetta} found that outside 2~au pre-perihelion, when outgassing rates were low, atomic and molecular emission features of 67P/Churyumov-Gerasimenko at UV/optical wavelengths were predominantly caused by dissociative electron impact excitation \citep{FeldAlice, Feldman2018, Bodewits2016, Galand2020}. Because gas production rates increased much faster than the ionizing solar radiation, a collisionopause was formed \citep{Nillson2017}. Within this boundary, collisions between water molecules and lower-energy electrons (below the threshold for dissociative impact excitation reactions) become dominant. For example, the formation of excited OH (A\ $^2\Sigma^+$) by electron impact on H$_2$O has an electron energy threshold of 9.1~eV, H Ly-$\alpha$ has a threshold of 15.4~eV, and the production of \ion{O}{1} 130.4 nm emission requires electron temperatures above 23.5~eV \citep{Beenakker1974, Bodewits2019}. At electron temperatures below 9~eV, collisions result in rovibrational excitation rather than dissociation of H$_2$O \citep{Itikawa2005}, and energy is radiated out of the inner coma in the form of IR emission.

\emph{Rosetta/Alice} recorded the emission of mostly atomic fragments in the FUV, including atomic hydrogen, carbon, and oxygen. It was noted that the relative intensities of lines (i.e., line ratios) were different from FUV spectra of previous comet observations \citep{FeldAlice}, which could not resolve the inner coma. The largest surprise was the OI] line at 135.6 nm, which is a forbidden intercombination multiplet that is rarely seen in coma spectra. These line ratios indicated that the dominant excitation process was dissociative electron impact excitation, rather than photofluorescence. In addition, the relative strength of the electron impact induced emission lines of \ion{H}{1}, \ion{O}{1}, and \ion{C}{1} in the FUV are highly sensitive to the parent molecule they are produced from, such as H$_2$O, O$_2$, and CO$_2$ (see Figure~\ref{fig:electron}). Electron impact emission thus provides a sensitive diagnostic probe of the local plasma environment in the coma and of the chemical composition of the atmospheres of small bodies. Remote studies using the \emph{Hubble Space Telescope} have used this to characterize the composition of the atmospheres of Europa \citep{Hall1995,Roth2014}, Callisto \citep{Cunningham2015}, and Ganymede \citep{Feldman2000, Roth2021}. \fix{However, remote studies using the \emph{Hubble Space Telescope} of 46P/Wirtanen found no evidence of dissociative electron impact in the inner coma  \citep{Noonan2021}, likely because the local densities were high enough to suppress them, as was observed by \emph{Rosetta} as 67P approached the Sun \citep{Bodewits2016, Galand2020}. Under what coma conditions dissociative electron impact can be the dominant process of emission remains an open question.} \\

\subsection{Solar Wind Charge Exchange }
\label{sec:charge_ex}
\begin{figure}[t]
\centering
\includegraphics[width=\columnwidth]{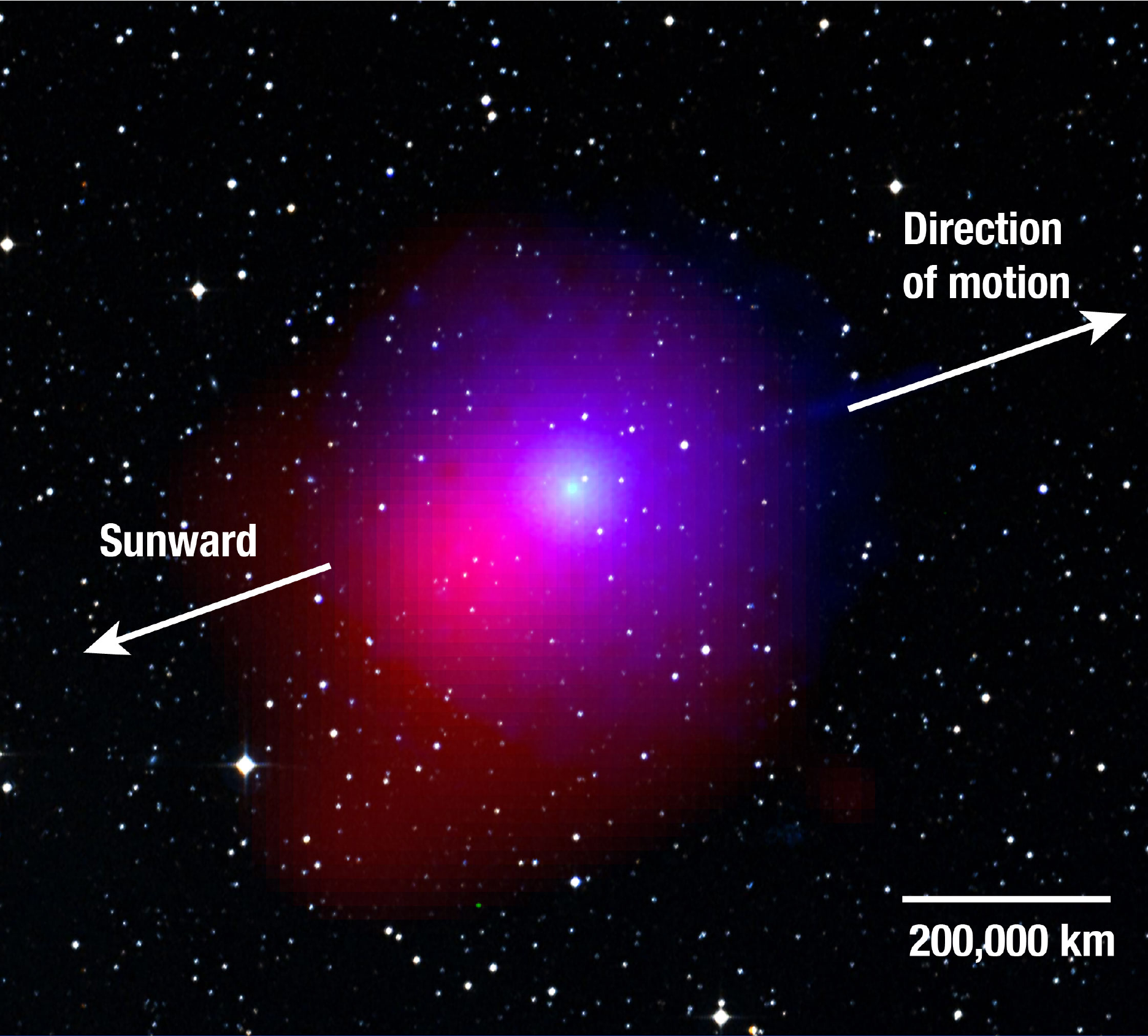}
\caption{Simultaneous observations of comet C/2007 N3 (Lulin) by the \emph{Neil Gehrels-Swift} observatory in soft X-rays (photon energies below 1~keV; red) and in the ultraviolet (200--350 nm; blue). In the ultraviolet, it sampled the fluorescent emission of OH radicals formed by the photodissocation of H$_2$O molecules. X-rays are emitted by solar wind ions coming from the direction of the Sun (bottom left), that capture electrons as they penetrate the neutral coma. Because the coma is collisionally thick to these incoming ions, the X-ray emission is asymmetric and offset towards the Sun. \label{fig:cx}}
\end{figure}

Comets emit up to 1~GW in extreme ultraviolet and X-ray radiation \citep{Lisse2004, Kras1997a}, despite the relatively low temperature of the gas in the coma (tens of Kelvins; Figure~\ref{fig:cx}). This emission is produced mostly by charge exchange between heavy ions in the solar wind (e.g., He$^{2+}$, O$^{7,8+}$, C$^{5,6+}$) and neutral molecules in the cometary atmosphere \citep{Cravens1997,Krasnopolsky1998}. The incoming ions capture one or more electrons directly into a highly excited state, which results in X-ray emission when the ion decays to its ground state. 

In the energy regime accessible by current X-ray satellites (typically above 0.25 keV), cometary charge exchange spectra consist mostly of emission lines of hydrogen- and helium-like ions. In particular, they are characterized by strong dipole forbidden lines, such as the \ion{O}{7} features around 565 eV, as metastable states are populated through cascades originating from the highly excited states electrons are captured into.  

Different from most optical telescopes, detectors on board X-ray facilities such as \emph{Chandra}, XMM-\emph{Newton}, and the \emph{Neil Gehrels-Swift} Observatory simultaneously provide spectral, spatial, and temporal information of incoming photons that can be used to probe properties of both the comet and the solar wind. \fix{For example, X-ray spectra can be used to measure the solar wind elemental composition and the charge state distribution of its ions, and thus of the conditions of its source region on the Sun \citep{Schwadron2000, Bod2007}. Because charge exchange emission is a quasi-resonant process, the angular momentum distribution of the captured electrons depends strongly on the velocity of the incoming ion and on electron donor species \citep{Beiersdorfer2003}. Charge exchange models predict that the resulting spectrum depends on the neutral donor species.}

Since the discovery of charge exchange emission in comets, it has been detected in many different astrophysical environments where a hot gas collides with a cold, neutral gas, including Mars, galaxies, and supernova remnants \citep{Dennerl2010}. Charge exchange reactions with the dominant coma gases (such as H$_2$O, CO$_2$, and CO) have typical cross sections of approximately $10^{-15}$ cm$^2$. This implies that for moderately active comets (gas production rates of less than a few times $10^{28}$ molecules~s$^{-1}$), solar wind ions will fly through most of the coma unhindered, and the charge exchange emission will map the neutral coma and jets, independent of the species of neutral gas. In this collisionally thin scenario, variations in the X-ray brightness depend on the heavy ion flux and neutral gas production rate of the comet. For comets with higher production rates, more of the coma will become collisionally thick to charge exchange and when observed under a sufficiently large phase angle, the X-ray morphology will take on a characteristic crescent shape (Figure~\ref{fig:cx}), first seen around comet C/1996 B2 (Hyakutake) \citep{Lisse1996}. Because the solar wind ions that are capable of emitting X-rays are depleted before they get close to the nucleus, the X-ray luminosity becomes decoupled from the gas production rate \citep{Dennerl1997} but not from variations in the solar wind \citep{Bonamente2021}. XMM-\emph{Newton} imaging of the particularly active comet C/2001 WM$_1$ (LINEAR) allowed for the first remote characterization of a comet's plasma bowshock, which is formed when the solar wind is decelerated by picking up heavy cometary ions \citep{Wegmann2005}.

X-ray astronomers are highly anticipating the launch of calorimeter instruments, which will provide a spectral resolution between 2--5~eV and good sensitivity to low energies. These upcoming observations will challenge our current understanding of charge exchange processes. For example, modelling suggests clear differences between spectra resulting from water group molecules and CO and CO$_2$ \citep{Mullen2017}. The spectral resolution of these instruments will allow testing of whether other plasma processes, such as electron bremsstrahlung contribute to cometary X-ray emission \citep{Krasnopolsky1998}. \fix{Finally, these observations will allow using comets as natural laboratories and address the question whether the state selective population by charge exchange results favors triplet or singlet states of Helium-like ions \citep{Mullen2017}. This question is very hard to address in the laboratory owing to the long life times of the relevant metastable states, yet the resulting emission lines are among the strongest in astrophysical charge exchange spectra}

\section{\textbf{Future Directions}}
\label{sec:nextsection}

\begin{figure*}[t]
\centering
\includegraphics[width=6in]{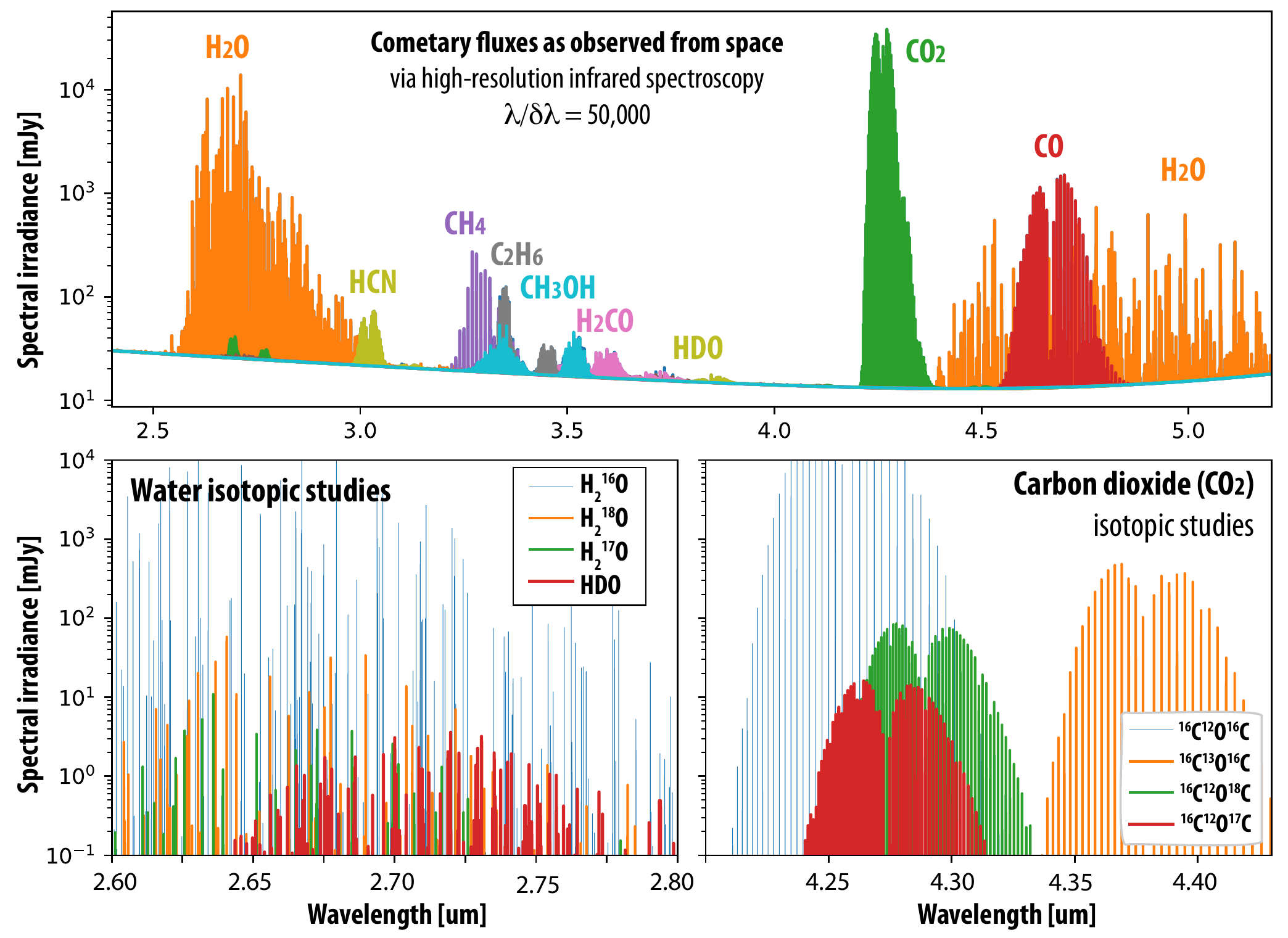}
\caption{ High-resolution spectroscopy of comets from space will enable probing of the organic, isotopic and isomeric composition of these primordial bodies with unique accuracy, in particular considering that many of these bright emissions are not available to ground-based observations due to telluric absorption. High-resolution spectroscopy also maximizes the contrast between the cometary gases and continuum levels, it reduces background noise, and it removes the spectral confusion impacting low-resolution studies (e.g., JWST). The simulations presented here are for a typical comet (5-10 per year) as observed with a 1m space observatory, and were calculated employing the Planetary Spectrum Generator \citep{psg2018}.\label{fig:space}}
\end{figure*}

\subsection{Spectro-Spatial Coma Observations}

The main observational breakthroughs in cometary exploration expected in the coming decades relate to the increased sensitivity and instantaneous bandwidth (across all wavelengths) afforded by improved instrumentation and larger telescope collecting areas. These developments will result in improved signal-to-noise and fidelity for \rev{atomic and molecular imaging and spectroscopy that will provide new insights into coma physics and chemistry through a better understanding of the gas macroscopic and microscopic motions and compositions.}

\fix{Most of modern research on cometary objects have been restricted to comets in the inner solar system, and typically to a limited number of species. This originates from the dramatic drop in sensitivity of comets with heliocentric distance $r_h$, and challenges in probing weak spectral signatures among many other bright primary emissions. For instance, fluorescence efficiencies scale by r$_h^{-2}$, global integrated flux densities scale by $\Delta^{-2}$ with distance, and if we consider insolation-driven activity, production rates would scale by r$_h^{-2}$. This would mean a dramatic dependence of r$_h^{-6}$ in the expected IR fluxes of distant objects, or r$_h^{-4}$ for radio/thermal emissions. Furthermore, the integration time needed for the same signal-to-noise ratio scales quadratically with the drop of flux in background dominated regimes (typically for $\lambda>3~\mu$m for ground-based observatories), and linearly in source's noise dominated regimes (typically at shorter wavelengths). Some trace species (such as Na, Ni, Fe, C$_2$H$_6$) have very strong signatures due to large fluorescence efficiencies, even though they have six orders of magnitude lower abundances than the primary volatiles.} Large collecting area observatories (e.g., E-ELT with 16 times greater collecting area than Keck and much improved in spatial sampling), space borne, cold observatories (e.g., JWST, with orders of magnitude improved sensitivity at thermal IR wavelengths), and in-situ probes are therefore essential to explore new frontiers in cometary research.

We anticipate the spectroscopic detection and characterization of new atomic and molecular species, as well as expanded data and statistics on species for which detailed spatial/spectral/large survey-sample information has thus far been unavailable. \rev{In the era of JWST}, the frontier of cometary studies will be expanded to fainter and more distant targets (r$_h>5$~au), particularly at IR wavelengths. This will allow the elucidation of novel physical and chemical regimes in the less well-explored cometary populations, such as Main Belt comets, active asteroids, Centaurs, Trans-Neptunian Objects, interstellar objects, and more distant/pristine Oort cloud comets for which volatile sublimation may not yet be fully activated \citep{Kelley2016}. 

Such advances will permit major discoveries and new insights regarding the fundamental chemical building blocks of comets. As we continue to explore deeper at every wavelength, new important markers are being identified, such as the recent discovery of ammonium salt absorption \citep{poch2020} and atomic metals fluorescing in the NUV/visible \citep{manfroid2021}. \fix{Future studies of coma isotope ratios have the potential to revolutionize our understanding of the origins and history of our solar system's most primitive materials \citep{Biver2023}}. More detailed mapping of coma density structures will help elucidate cometary activity and outgassing mechanisms, and expanded, more spatially-complete, higher-resolution surveys of coma rotational temperatures will provide new theoretical challenges towards a complete understanding of coma heating and cooling processes.

\fix{Increased spectral resolution will diminish spectral confusion, which is particularly severe in the optical/infrared, where many molecular and isotopic signatures overlap. Spectroscopic surveys at high spectral resolution ($\lambda / \delta\lambda > 50,000$) will permit to properly separate isomeric and isotopic signatures, and open new avenues for better understanding the excitation of fragment species}. For example, \citet{nel_coch2019} found that spectra of the C$_2$ Swan bands ($\lambda / \delta\lambda~\sim~60,000$) in two comets are best described by a bimodal rotational temperature distribution. The underlying mechanism can be tested by high-resolution optical studies of comets spanning a range in heliocentric and cometocentric distances. The authors point out that such studies can first test if two-temperature distributions are a common feature in fluorescence spectra of cometary C$_2$, and second, test hypotheses of the underlying mechanisms, for example competing formation pathways from the dissociation of parent molecules. 

\fix{\citet{Biver2023} emphasize the importance of spatially-resolved measurements in understanding the storage of volatiles in comet nuclei and their release in the coma. \citet{bonev2021} discuss how clues about associations (or differences) in volatile release deduced from ground-based spatial studies can be linked to the detailed findings from the Rosetta and EPOXI mission to comets. As discussed in this chapter, spatial distributions are also informative of various excitation coma processes, including the effects of icy grain vaporization on coma temperatures. The basic parameters that limit spatial studies are the spatial scale (arcsec/pix) and the dimension(s) of the field-of-view (arcseconds, respectively km in projected distance on the sky plane, the latter depending of geocentric distance $\Delta$). In the infrared, the optimal combination of detector sensitivity and long-slit capability ($\sim0.13$ arcsec/pix) is currently offered by NIRSPEC at Keck 2. The total slit length (24$''$) of NIRSPEC allows for significant extension in projected distance ($> \sim 2,000-3,000$ km at $\Delta$ = 1~au), as needed for detecting molecular emission farther from the comet nucleus, including from extended coma sources. A key challenge in optimizing the design and cost of future infrared high-resolution spectrographs may be incorporating increased spectral coverage, while maintaining slit length comparable to or exceeding that of NIRSPEC.}

\rev{Looking beyond present-day instrumentation}, a future space observatory offering high-resolution IR spectroscopy (at $\lambda / \delta\lambda~\gtrsim~50,000$, see Figure~\ref{fig:space}), would enable sampling of narrow cometary emission lines with unprecedented sensitivity and precision, therefore allowing measurement of many isotopic and spin species of H$_2$O and CO$_2$, CO and trace organic species. Future and planned telescopes (e.g., \emph{European Extremely Large Telescope}, \emph{Thirty Meter Telescope}, \emph{Giant Magellan Telescope}) with large collecting areas will permit exploration of ever fainter objects at increasingly large heliocentric distances and tracking cometary activity as they enter the inner Solar System. Integral Field Unit (IFU) instrumentation with 2D NUV/Optical/IR spatial-spectral mapping capabilities would allow identification of heterogeneous regions of activity in the nucleus and secondary sources in the coma, as well as mapping of inner coma temperatures in a large number of comets with a wide range of production rates.

At longer wavelengths, multi-beam radio/mm/sub-mm receiver technology has opened up the field of high spectral-resolution coma mapping, enabling snapshot molecular imaging in a small fraction of the time it takes to perform a conventional raster-style map using a single-pixel receiver \citep{cou17,cou20}. Inclusion of larger focal plane receiver arrays in future generations of mm/sub-mm telescopes will facilitate more detailed spatial studies of coma physical and chemical properties. High-sensitivity, spatially complete coma maps are difficult to obtain with current telescopes, even with ALMA, due to the interferometer's inherent lack of sensitivity to extended structures. Therefore, science related to the mapping of less abundant molecules (such as isotopologues or complex organics) and to the determination of kinematics and production/release mechanisms will benefit strongly from next-generation IFU-style heterodyne focal plane arrays. Improved sensitivity for molecular mapping will be ensured thanks to ongoing (and planned) upgrades to currently-available radio interferometers.

\subsection{Fundamental Spectroscopic Data Requirements}

The availability of sufficiently accurate theoretical and laboratory data limits the accuracy with which astronomical observations can be interpreted. New observational discoveries and improved spectroscopic sensitivity therefore drives the need for improved gas-phase experimental and spectroscopic data.

The capabilities of state-of-the-art radio interferometers, such as ALMA, have made it possible to probe a broader range of spatial scales than previously possible via radio/sub-millimeter molecular emissions (particularly in the inner coma). Accurate interpretation of rotational emission lines originating from such a broad range of nucleocentric radii places strict demands on the reliability of the underlying physical model. Major uncertainties still remain in our knowledge of state-to-state collisional excitation/de-excitation rate coefficients ($C_{ij}$), applicable to the calculation of rotational excitation of coma gases (Section \ref{sec:col_rates}). More accurate collision rates between less abundant molecules and water $C_{ij}$(X--H$_2$O), where the impactor $X$ can be HDO, CH$_3$OH, H$_2$CO, HNC, HC$_3$N, CH$_3$CN, NH$_2$CHO and CS), are required. Future studies will also benefit from $C_{ij}$ data for molecules colliding with CO and CO$_2$ (for application to comets farther from the Sun where H$_2$O may not be the dominant collision partner), as well as with electrons \citep{Ahmed2021}. To realize the full potential of such detailed, non-LTE calculations, it will be necessary to track the rotational temperatures of all molecular collision partners as they depart from thermal equilibrium, since in general, the $C_{ij}$ values depend on the rotational state of the impactor (characterized by $T_{rot}$), as well as the impact velocity (from $T_{kin}$).

The high spatial resolutions expected at shorter wavelengths owing to future extreme adaptive optics and ELTs will pose strong challenges to the interpretation of the resulting ultraviolet / optical / infrared emission. Modeling the coma distributions of emitting molecules requires knowledge of the photodissociation lifetimes in the solar radiation field, which are still not well known for many molecules. Understanding the spatial distributions of emitters will also require the incorporation of heterogeneous outgassing, accounting for multiple native and extended sources of major species (H$_2$O, CO$_2$, CO) and trace species (C$_2$H$_2$, HCN, etc.). It is not yet clear what is the excitation state (or how it is being affected) of molecules released from different extended sources such as dust, from icy grains, or in some photochemical reactions.

Similarly, many electron impact emission features observed by \emph{Rosetta/Alice} and \emph{Rosetta/OSIRIS} at 67P/Chury\-umov-Gera\-simenko have only been measured at a fixed impact energy, but not between their threshold energy and approximately 100~eV, the energy range that includes the majority of electrons observed at the comet. In the case of solar wind charge exchange, laboratory data are needed for reactions with the main relevant molecules. Such data include multi-electron capture processes and velocity-dependent charge-exchange cross sections.

As discussed in the previous sections, the completeness and accuracy of spectroscopic line lists play a critical role in the calculation of model fluxes for the processes discussed in this chapter. In the last decade, there has been a dramatic increase in the availability of high-temperature/energy ab-initio line lists, yet there is still a lack of databases addressing species that are prevalent in cometary atmospheres but less common in planetary atmospheres (e.g., C$_2$H$_6$, CH$_3$OH, various cations such as H$_2$O$^+$, CO$^+$; \citealt{Fortenberry2021}). Such line lists with detailed transition level identifications (needed for fluorescence calculations) and high-energy range will potentially allow identification of the tens of thousands of currently unassigned lines \citep{Cochran2002,Dello2016}, enable sensitive probes of isotopic and isomeric ratios, and allow to better characterize prompt emission efficiencies (including their variations driven by variable solar activity) with proper assessment of dissociation yields in photochemistry (e.g., excited oxygen atoms from the photodissociation of CO$_2$ vs. H$_2$O).

\subsection{\rev{Comprehensive Probing of Cometary Processes}}

\rev{In the last decade, our capabilities to characterize the structure and composition of comets and the processes active in the coma have grown tremendously. In-situ probing via the \emph{Rosetta} mission (e.g., mass spectrometry, inner coma remote sensing) and coordinated near-simultaneous observations across wavelengths have become increasingly common (e.g., \citealt{Jehin2009, Meech2011}).} Observations that bridge multiple wavelength regimes and techniques not only benefit compositional studies but are also critical to exploration of novel excitation processes. For instance, coordinated IR and radio observations of parent species (e.g., HCN, CH$_3$OH, H$_2$CO) enable a deeper investigation of the origins of these species, which may be both precursor molecules or dissociation fragments themselves. Furthermore, coordinated optical/IR/radio observations better elucidate parent-fragment relationships, such as C$_2$H$_2$ -- C$_2$, HCN -- CN, NH$_3$ -- NH$_2$ -- NH. \rev{Similarly, the derived kinematics, composition, and isotopic ratios from in-situ mass spectrometry provide new challenges to the interpretation of (and consistency with) remote sensing observations.} For instance, multi-wavelength observations can resolve the disagreements in the  retrieved columns of species at different wavelengths, and provide a better understanding of the overall excitation processes and the impact of new modeling/analysis methods \citep{lippi2021}. A primary example of this is the long-standing disagreement in the production rates of hydrogen cyanide and other species when retrieved through IR and radio techniques \citealt{villanueva2013}. The production rates of HCN, determined from rovibrational transitions (IR fluorescence), often exceed those obtained from observations of pure rotational transitions, by about a factor of two. Evaluating the role of cascades in IR fluorescence (Section \ref{sec:fluoresce}) did not bring results closer, so the challenge remains to critically examine the assumptions regarding HCN excitation made by by each technique, and the physical and chemical processes involved in the storage and release of cometary volatiles, together with their subsequent processing in the coma.

Decades-long observations of the CN red and violet band systems have been important in compositional classifications of comets \citep{AHearn1995, Fink2009}. It is of note that the source of CN is not entirely clear \citep{Fray2005, altwegg2020}. Photodissociative excitation of HCN does not appear to be a major source of electronically excited CN \citep{Bock85}, but this might be different for other possibe sources of this molecule (\emph{e.g.} parent molecules, dust). \citet{pagamumma2016} recently proposed a new excitation model for CN fluorescence and tested it against high-resolution IR spectra of comet C/2014 Q2 (Lovejoy). This model includes both electronic transitions in the $A~^{2}\Pi$ -- $X~^{2}\Sigma^{+}$ and $B~^{2}\Sigma^{+}$ -- $X~^{2}\Sigma^{+}$ systems and rovibrational transitions within the CN ground electronic state. The IR can therefore sample both CN and its likely precursor HCN. Furthermore, contemporaneous IR and optical observations of CN at high spectral resolution are now feasible with facilities that allow flexible scheduling, like \emph{McDonald Observatory} and {\sc nasa}'s {\sc irtf}. Such observations would facilitate intercomparison of the assumptions and parameters of existing models for CN fluorescence, thereby allowing the excitation of this frequently observed radical to be traced from the inner collisional coma (IR) to the extended coma (optical). 

\rev{Integrating analyses across wavelength regimes, with localized in-situ mass-spectrometry measurements and magneto-hydrodynamic probing, would further advance ongoing studies of non-LTE and dissociative excitation in cometary comae}. For example, there are systematic uncertainties in the release rates from a particular precursor (CO$_2$ or H$_2$O) leading to metastable atomic states (O\ $^1$D and O\ $^1$S) whose long lifetimes present a challenge to laboratory measurements (Section~\ref{sec:prompt}). \citet{mckay2015} demonstrated that comparisons of CO$_2$ production rates from \ion{O}{1} (ground-based optical observations) and from direct CO$_2$ detections (IR observations from space) can constrain dissociation release rates empirically.

\section{Outlook}

Current spectroscopic studies have, out of necessity, focused on the brightest comets. The increased detector sensitivity of spectrographs creates opportunities for remote, molecular detections in comets that could not be previously observed. This will open in-depth exploration of new radiation and density regimes. Interpreting volatile emissions from the extremely low-density environments of comets with marginal activity, \fix{or with activity driven by volatiles other than H$_2$O,} will require re-evaluating the balance between collisional and radiative excitation processes. Such regimes are expected in distant comets. In addition, \citet{nuth2020} suggested that active comets, volatile-rich active asteroids, and `dormant' small bodies in the solar system represent a single evolutionary sequence, motivating the need to bridge radiative diagnostics of these objects, including intrinsically weak comets observed closer to the Sun.

\vskip .5in
\noindent \textbf{Acknowledgments.} \\

D. B. acknowledges support from NASA as part of the Rosetta Data Analysis Program (No.80NSSC19K1304). M. A. C. acknowledges support from the National Science Foundation (Grant No. AST-2009253) and the NASA Planetary Science Division Internal Scientist Funding Program, through the Fundamental Laboratory Research work package (FLaRe). G. L. V. acknowledges support from NASA as part of the US participation in the Comet Interceptor mission. B. P. B. acknowledges support from the National Science Foundation (Grant No. AST-2009398) and NASA Solar System Workings Program (Grant No. 80NSSC20K0651).

\bibliographystyle{sss-full_new.bst}
\bibliography{refs.bib}

\end{document}